\documentclass[namedate]{article}

\PassOptionsToPackage{numbers, compress}{natbib}
\usepackage[final]{arxiv_pkg}

\usepackage[utf8]{inputenc}      
\usepackage[T1]{fontenc}         
\usepackage[
	hidelinks,
	colorlinks=true,
	linkcolor=blue,
	citecolor=blue,
	urlcolor=blue
]{hyperref} 
\usepackage{url}                 
\usepackage{booktabs}            
\usepackage{amsfonts}            
\usepackage{amsmath}
\usepackage{nicefrac}            
\usepackage{microtype}           
\usepackage{subfiles}
\usepackage{xcolor}
\usepackage{multirow}
\usepackage{enumerate}
\usepackage{subfiles}
\usepackage{array}
\usepackage{multirow}
\usepackage{threeparttable}
\usepackage{graphicx}
\usepackage{subfiles}
\usepackage{orcidlink}
\usepackage{algorithm}
\usepackage{algpseudocode}
\usepackage{chngcntr}
\usepackage{booktabs}
\usepackage{tabularx}
\usepackage{placeins}

\graphicspath{{Figures/}{Supplemental_Figures/}}

\title{\Large ViSAR: Training-Free Adaptive-$k$ Retrieval for\\Document Visual Question Answering}

\makeatletter
\renewcommand{\@fnsymbol}[1]{\@arabic{#1}}
\makeatother

\author{
	\AND
	Adrien Mialland \\
	INSA Lyon, CNRS, LIRIS UMR 5205, F-69621 Villeurbanne, France\\
	\texttt{mialland.a@gmail.com} \orcidlink{0000-0001-5359-674X}
	\And
	Marc Plantevit\\
	EPITA Research Laboratory (LRE), FR-94276, Le Kremlin-Bicêtre, France\\
	\texttt{marc.plantevit@epita.fr}
	\And  
	Julien Gallois\\
	Lowit, FR-69003 Lyon, France\\
	\texttt{julien.gallois@lowit.fr}
	\And
	Céline Robardet \\
	INSA Lyon, CNRS, LIRIS UMR 5205, F-69621 Villeurbanne, France\\
	\texttt{celine.robardet@insa-lyon.fr}
}

\newcommand{\GithubViSAR}{https://github.com/adrienmialland/ViSAR}
\providecommand{\keywords}[1]
{
	\small	
	\textbf{\textit{Keywords---}} #1
}

\makeatletter
\newcommand{\figurecaption}[1]{%
	\begingroup
	\def\@captype{figure}%
	\caption{#1}%
	\endgroup
}
\makeatother

\makeatletter
\newcommand{\tablecaption}[1]{%
	\begingroup
	\def\@captype{table}%
	\caption{#1}%
	\endgroup
}
\makeatother

\definecolor{stagecolor}{RGB}{150,35,45}

\begin{document}
	
	\maketitle
	
	\begin{abstract}
		Document Visual Question Answering (DocVQA) often leverages Retrieval-Augmented Generation (RAG), where late-interaction encoders are commonly used to identify document pages relevant to a user query, before answer generation by a Large Vision-Language Model (LVLM). Existing approaches typically retrieve a fixed top-$k$ number of pages regardless of query complexity, which increases LVLM latency and may degrade answer accuracy when fewer pages would be enough. We introduce ViSAR (Visual Semantic Activation Retrieval), a training-free, budget-aware adaptive-$k$ retrieval method for late-interaction visual document retrieval. ViSAR operates directly in the embedding space to construct a query-conditioned page-level similarity matrix that highlights query-relevant semantics and dynamically determines the number of pages to retrieve, within a given page budget. We evaluate our method across multiple datasets, encoders, and LVLMs. ViSAR retrieves compact, query-adapted page sets that reduce end-to-end RAG latency by up to 58.9\%, with no statistically significant accuracy loss and significant gains in several settings, compared with fixed top-$k$ and adaptive retrieval heuristics. Furthermore, we show that the similarity matrix structure correlates with answer accuracy, suggesting future directions for retrieval quality-aware document understanding. Our code is available at \url{\GithubViSAR}.
	\end{abstract}
	

	\section{Introduction}
	
	\begin{figure}[t!]
		\centering
		\includegraphics[width=1\columnwidth]{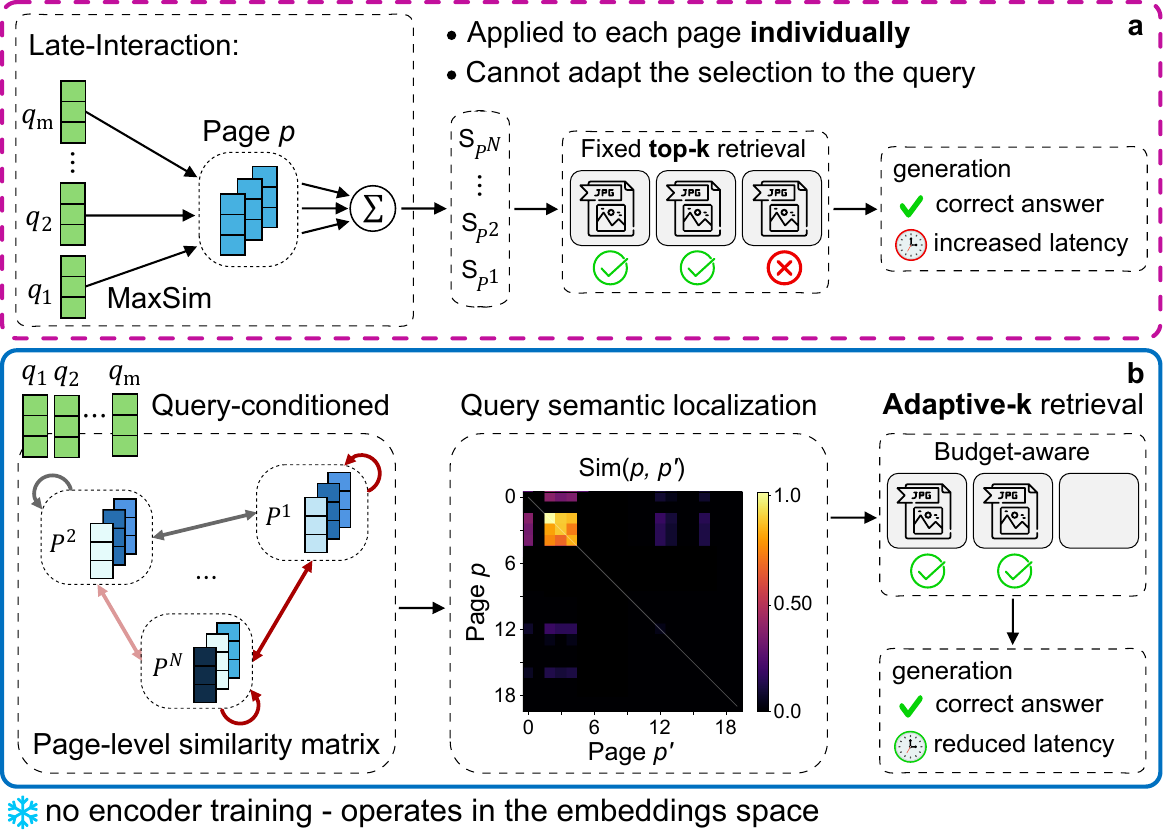}
		\vspace{-10pt}
		\caption{Document page retrieval mechanisms. (a) Late-interaction enables fine-grained query-page matching using multi-vector representations, but it relies on fixed top-$k$ retrieval that cannot adapt to the query, introducing irrelevant pages and unnecessary latency. (b) The proposed method, ViSAR, leverages the same multi-vector representations, without additional training, to build a page-level similarity matrix and enable adaptive-$k$ retrieval, reducing irrelevant pages and lowering latency. ViSAR is budget-aware (represented by the light grey boxes), only computing the required similarities to select an adaptive number of pages that fit within the budget, without approximation.}
		\label{fig:ViSAR_vs_SOTA} 
	\end{figure}
	
	Document Visual Question Answering (DocVQA) aims to answer a user query about visually rich documents using page screenshots, and has attracted growing interest across many applications \cite{M3DocRAG,mdocagent,visrag,hierarchicamDocVQA,screenAI}. These documents often include text, images, charts, or tables that blend into document-specific layouts, requiring methods capable of jointly understanding the query and heterogeneous visual content to localize relevant information and generate an answer. DocVQA is therefore commonly addressed using Large Vision-Language Models (LVLMs) \cite{Qwen3vl,InternVL,surveyVRDOCVQA}, which reason over textual and visual modalities jointly. Recent progress has demonstrated close-to-human performance in single-page DocVQA \cite{DocVQAdataset,ChartVQA,InfoVQA}, but processing multi-page documents is more costly, as each additional page substantially increases the LVLM latency, while most pages contain no information relevant to the query, which makes it harder to gather the relevant evidence. Handling long documents therefore requires selecting the relevant subset of pages.
	
	In this setting, Retrieval-Augmented Generation (RAG) emerges as a promising paradigm, where an encoder projects a query and the document pages into a shared embedding space, and a retrieval step selects the most relevant pages for an LVLM to generate an answer. However, determining how many pages should be retrieved remains an open challenge, despite its critical impact: too many pages increase LVLM latency and may introduce irrelevant context that degrades answer accuracy, whereas too few pages risk omitting relevant evidence. In practice, the retrieval step selects a fixed top-$k$ number of pages for every query, paying the full latency cost even when fewer pages would be enough.
	
	State-of-the-art retrieval methods rely on a late-interaction mechanism \cite{colbert}, where the encoder independently maps the query and the pages into multi-vector representations. Page relevance is obtained by aggregating the best matches of each query embedding across page embeddings, enabling fine-grained semantic matching, offline page encoding, and scalable online retrieval. However, while late-interaction models have significantly improved page representation \cite{colbert,Colpali,ModernVBert,visrag}, they produce independent page relevance scores and are still used with fixed top-$k$ retrieval, which adds latency and noisy LVLM input context as it cannot adapt the number of pages to the query. Moreover, by treating pages independently and assigning uniform importance to embeddings, late-interaction does not exploit the semantic structure across pages (\autoref{fig:ViSAR_vs_SOTA}-a).
	
	Text retrieval methods have mitigated this through embedding weighting based on token frequency statistics \cite{colbertIDF,TWBERT,TokenImportance,SLIM}, or adaptive-$k$ retrieval \cite{AdaptiveK}, but they rely on discrete token structures that do not extend to visual embeddings and often require additional training. In contrast, we show that late-interaction representations reveal sparse, query-dependent semantic content across pages, enabling dynamic localization and adaptive selection of the number of pages to retrieve.
	
	We introduce ViSAR (Visual Semantic Activation Retrieval), a training-free retrieval mechanism that operates directly in the embedding space of late-interaction encoders to adapt the number of pages. ViSAR constructs a query-conditioned page-level similarity matrix that highlights query semantics and enables adaptive-$k$ retrieval, dynamically determining the number of pages within a given page budget (\autoref{fig:ViSAR_vs_SOTA}-b). By providing fewer pages to the LVLM, ViSAR reduces end-to-end RAG latency while preserving answer accuracy. We also show that the structure of the similarity matrix correlates with answer accuracy, suggesting a general principle for quality-aware document understanding. Our main contributions are as follows:
	
	\begin{itemize}
		\item \textbf{ViSAR:} A training-free, budget-aware adaptive-$k$ retrieval mechanism for visual retrieval using late-interaction encoders, which adapts the LVLM context to each query.
		\item \textbf{Latency reduction:} By retrieving fewer pages on average than fixed top-$k$ retrieval, ViSAR reduces end-to-end RAG latency by up to 58.9\%. Across three encoders and five LVLMs, it shows no statistically significant accuracy loss, and significant gains in several settings.
		\item \textbf{Similarity structure:} A query-conditioned page-level similarity matrix whose structure reflects evidence localization and correlates with answer accuracy.
	\end{itemize}
	
	\section{Related Work}
	
	\subsection{Multi-Vector Retrieval and Embedding Weighting.} Multi-vector retrieval via late-interaction was introduced by ColBERT \cite{colbert} for text documents. It represents queries and pages as independent sets of token-level embeddings, and page relevance is computed by aggregating query embeddings' best similarity score to page embeddings. Subsequent works improved its efficiency \cite{colbertv2,PLAID}, introduced token weighting mechanisms based on frequency statistics, learned importance estimation, or sparse representations \cite{colbertIDF,TokenImportance,TWBERT,SLIM,SPLADEv2,ContextAwareImportance,TopicEnrichedEmbedings}, and explored retrieval refinement through pseudo-relevance feedback \cite{ColbertPRF}. However, these approaches rely on Optical Character Recognition (OCR) to extract the text content, cannot capture visual content, and often require additional training. Recent works such as ColPali \cite{Colpali,ModernVBert} extend late-interaction by encoding each page as a set of visual patch embeddings, emerging as the state-of-the-art paradigm for OCR-free visual document retrieval \cite{ViDoReV3,Colpali}. However, it still relies on independent page scores and fixed top-$k$ retrieval, leaving the semantic organisation of pages unexplored. ViSAR instead exploits this structure to enable adaptive-$k$ retrieval.
	
	\subsection{Adaptive-$k$ Retrieval.} Adaptive retrieval has mainly been explored via iterative approaches, where LLMs or LVLMs reason over multiple rounds of fixed top-$k$ retrieval, making $k$ vary implicitly across iterations \cite{React,SelfRAG,AdaptiveRAG,SimpleDoc,DocReact}. However, adaptive-$k$ methods directly estimate the number of pages to retrieve in a single pass. Prior work determines the retrieval cut-off via heuristics applied to late-interaction scores by identifying the largest gap between consecutive scores \cite{AdaptiveK} or by clustering scores to detect transitions \cite{CulsterAdaptiveK,AVIR}. However, these methods are text-based or require additional training and rely solely on score distributions, while ViSAR performs training-free adaptive-$k$ retrieval by exploiting the semantic structure encoded in late-interaction representations.
	
	\subsection{Visual Document Retrieval and DocVQA.} These advances in retrieval have directly influenced recent OCR-free DocVQA systems, where effective page selection is a key component of the reasoning pipeline combined with LVLM-based reasoning \cite{RAGVisionSurvey}. VisRAG \cite{visrag} introduced a dedicated single-vector visual retriever, while M3DocRAG \cite{M3DocRAG} demonstrated the effectiveness of multi-vector late-interaction retrieval with ColPali. Other frameworks \cite{VDocRAG,mdocagent,VisDoMRAG,MHierRAG} improve visual encoders, iterative retrieval, or LVLM reasoning. However, none exploit the semantic structure induced by late-interaction representations, which ViSAR addresses while remaining agnostic to the encoder and LVLM.

	\section{ViSAR: Visual Semantic Activation Retrieval}
	
	\begin{figure}[t!]
		\centering
		\includegraphics[width=\linewidth]{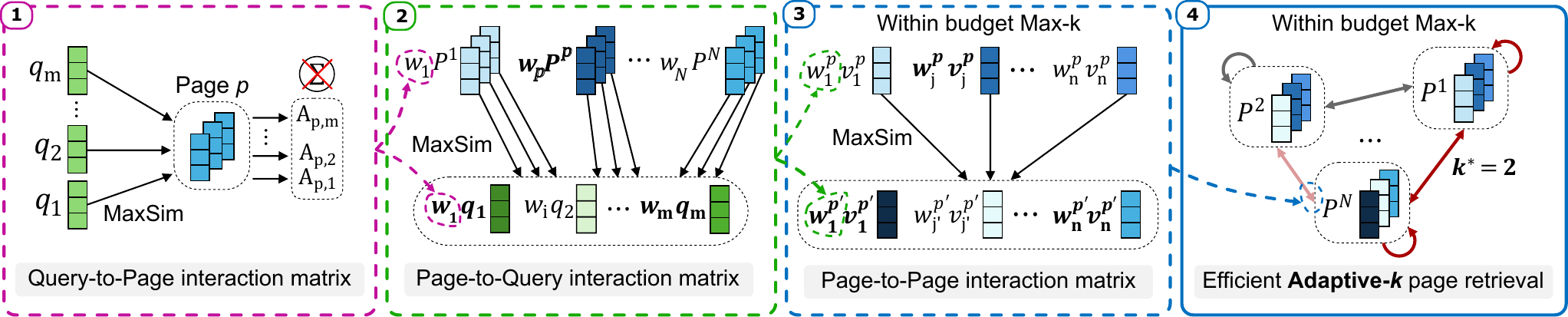}
		\vspace{-10pt}
		\caption{ViSAR overview. The method performs successive interactions in embedding space. \textbf{1. Query-to-page:} Unlike late-interaction, MaxSim activations are analyzed across pages rather than summed over tokens, to derive token- and page-level weights (pink dotted arrows). \textbf{2. Page-to-Query:} These weights modulate patch-query similarities to derive patch-level weights (green dotted arrows). \textbf{3. Page-to-Page:} These weights modulate patch-patch similarities to compute query-conditioned similarities (blue dotted arrow), forming a page-level similarity matrix that highlights the shared query-related content. \textbf{4. Adaptive-$k$:} The similarity matrix is used to retrieve an adaptive-$k$ set of pages. The pages are selected under a Max-$k$ budget, which computes only the required similarities to select at most Max$k$ pages, without approximation, which limits both steps 3 and 4 computation. ViSAR thus adds marginal latency while substantially reducing the downstream answer generation latency, by retrieving fewer pages on average than fixed top-$k$ retrieval (\autoref{fig:ViSAR_vs_SOTA}).}
		\label{fig:ViSAR_schema}
	\end{figure}
	
	\autoref{fig:ViSAR_schema} depicts our method, ViSAR, which addresses visual retrieval in DocVQA. ViSAR enables adaptive-$k$ retrieval by weighting Query-to-Page, Page-to-Query, and Page-to-Page interactions without encoder retraining. It uses the MaxSim operator from late-interaction \cite{colbert,Colpali} to exploit the fine-grained semantic structure in the embedding space and improve retrieval. Algorithm \autoref{alg:visar} summarizes the full pipeline.
	
	\subsection{Preliminaries: MaxSim and Late-Interaction}
	
	We denote a document $\mathcal{D}$ as a set of pages $\mathcal{D} = \{ P^{1}, \dots, P^{N} \}$. Each page $P^p$ and query $Q$ are denoted as $P^{p} = \{ v^{p}_j \}_{j=1}^{n_p}$ and $Q = \{q_i\}_{i=1}^{m}$, their respective sets of multi-vector representations in a common embedding space $\mathbb{R}^D$, such that $P^{p} \in \mathbb{R}^{n_p \times D}$ and $Q \in \mathbb{R}^{m \times D}$. The relevance score of $P^{p}$ to $Q$ is estimated via late-interaction between their sets of embeddings (\autoref{eq:colpali}), summing each $q_i$ maximum similarity over all $v^{p}_j$, where $\mathrm{sim} \!\left( \cdot, \cdot \right) = \langle \cdot, \cdot \rangle$ defines the cosine similarity between two embeddings.
	\begin{equation}
		\label{eq:colpali}
		S_{Q, P^p} = \sum^m_{i=1} \max_{j \in \{1, \dots, n_p\}} \mathrm{sim} \!\left( q_i,\; v^{p}_j \right)
	\end{equation}
	
	\subsection{Adaptive Multi-Level Interaction Weighting}
	
	Given the uniform contribution of $q_i$ and $v^{p}_j$ in \autoref{eq:colpali}, we estimate their importance directly from the embedding space. Since cosine similarity measures embedding alignments in a semantically structured space, the MaxSim operator over all $v^{p}_j$ captures how strongly the semantic encoded by $q_i$ is realized in $P^p$. We refer to $A_{p,i}$ (\autoref{eq:activation_qi}) as the \emph{activation} score. Unlike late-interaction (\autoref{eq:colpali}), which aggregates the activations over $q_i$ into a single score, we exploit the full activation matrix to identify discriminative query semantics.
	\begin{equation}
		\label{eq:activation_qi}
		A_{p,i} = \max_{j \in \{1, \dots, n_p\}} \mathrm{sim} \!\left( q_i,\; v^{p}_j \right) = \max_{j \in \{1, \dots, n_p\}} \langle q_i,\; v^{p}_j \rangle
	\end{equation}
	
	\subsubsection{Query-to-Page Interaction Weighting} We use $A_{p,i}$ to derive weights for vectors $q_i$ and pages $P^p$ to highlight strong and spatially localized semantics. $A_{p,i}$ is rescaled by its mean across pages into $\hat{A}_{p,i}$, and modulated by its normalized inter-page standard deviation $\hat{\sigma}_i$, yielding $\tilde{A}_{p,i}$ (\autoref{eq:relative_activation}).
	\begin{equation}
		\hat{A}_{p,i} = \frac{A_{p,i}} {\mathrm{mean}_{p} \!\left( A_{p,i} \right)}
	\end{equation}
	
	\begin{equation}
		\hat{\sigma}_i = \frac{\mathrm{std}_p \!\left( A_{p,i} \right)} {\mathrm{mean}_{i'} \!\left( \mathrm{std}_p \!\left( A_{p,i'} \right) \right)}
	\end{equation}
	
	\begin{equation}
		\label{eq:relative_activation}
		\tilde{A}_{p,i} = \hat{A}_{p,i} \cdot \hat{\sigma}_i
	\end{equation}
	Aggregating $\tilde{A}_{p,i}$ over pages penalizes ubiquitous semantic content and highlights sparse activations, resulting in a weight $w_i$ for each $q_i$ (\autoref{eq:w_i}).
	\begin{equation}
		\label{eq:w_i}\textbf{}
		w_{i} = \log\frac{N}{1 + a_{i}},
		\quad\text{with }
		a_{i} = \sum_{p} \tilde{A}_{p,i}
	\end{equation}
	We then compute a page-level semantic co-activation $C_{i,i'}^{p}$, which is normalized by the mean activation of each $q_i$ across pages, and a weight $w_p$ is obtained for each page by aggregating the mean co-activation of all $q_i$ (\autoref{eq:w_p}).
	\begin{equation}
		C_{i,i'}^{p} = \tilde{A}_{p,i} \cdot \tilde{A}_{p,i'}
		\qquad
		\bar{A}_i = \mathrm{mean}_{p} \big(\tilde{A}_{p,i}\big)
	\end{equation}
	
	\begin{equation}
		\label{eq:w_p}
		w_{p} = \sum^m_{i=1} \mathrm{mean}_{i'} \Big(\frac{C_{i,i'}^{p}}{\bar{A}_i \cdot \bar{A}_{i'}}\Big)
	\end{equation}
	
	\subsubsection{Page-to-Query Interaction Weighting} Min-Max normalization is applied to $w_{i}$ and $w_{p}$, yielding $\hat{w}_{i}, \hat{w}_{p} \in [0,1]$, which modulate the activations $\smash{\tilde{A}_{p,i}}$ and weight the cosine similarity between all $q_i$ and $v^{p}_j$. The MaxSim operator is then applied over all $q_i$ (i.e., \emph{patch-to-query direction, reversing standard late-interaction}) to capture how strongly the semantic encoded by $v^{p}_j$ is realized in the query $Q$. We use this weighted maximum similarity as the relevance score $r^{p}_{j}$ of $v^{p}_j$. A weight $w^{p}_{j}$ is obtained for each $v^{p}_j$ of a page $P^p$ by centering and thresholding the relevance $r^{p}_{j}$ (\autoref{eq:w_pj}).
	\begin{equation}
		\label{eq:activation_vpj}
		r^{p}_{j} = \max_{i \in \{1, \dots, m\}} \Big[ \langle v^{p}_j,\; q_i \rangle \cdot \big( \tilde{A}_{p,i} \cdot \hat{w}_{i} \cdot \hat{w}_{p} \big)^2 \Big]
	\end{equation}
	
	\begin{equation}
		\label{eq:w_pj} 
		w^{p}_{j} = \max\!\left( 0,\; r^{p}_{j} - \mathrm{mean}_{p,j}\!\left( r^{p}_{j} \right) \right)
	\end{equation}
	
	\subsubsection{Page-to-Page Interaction Weighting} A Min-Max normalization is applied to $w^{p}_{j}$, yielding $\hat{w}^{p}_{j} \in [0, 1]$, which modulates the importance of page embeddings and enables a similarity measure between pages. For a source page $P^p$ and target page $P^{p'}$, the cosine similarity between all $v^{p}_j$ and $\smash{v^{p'}_{j'}}$ is weighted by the target embedding weights $\hat{w}^{p'}_{j'}$. The MaxSim operator is then applied over all $\smash{v^{p'}_{j'}}$ to capture how strongly the semantic encoded by $v^{p}_j$ is realized in $\smash{P^{p'}}$. The resulting score is then weighted by the source embedding weight $\hat{w}^{p}_{j}$ to obtain $S^{p \rightarrow p'}_{j}$ (\autoref{eq:sim_ppj}).
	
	\begin{equation}
		\label{eq:sim_ppj}
		S^{p \rightarrow p'}_{j} = \hat{w}^{p}_{j} \cdot \max_{j'} \Big[ \langle v^{p}_j,\; v^{p'}_{j'} \rangle \cdot \hat{w}^{p'}_{j'} \Big]
	\end{equation}
	
	Finally, the similarity matrix $\mathrm{Sim}(p,p') \in \mathbb{R}^{N \times N}$ is computed by averaging the $T$ largest interactions of $S^{p \rightarrow p'}_{j}$ across source embeddings $v^p_j$, denoted by $\mathcal{T} \subset \{1,\ldots,n_p\}$, and taking the square root (\autoref{eq:sim_matrix}). Since each source embedding independently searches for its best match in the target page (\autoref{eq:sim_ppj}), this formulation naturally produces a directional page similarity, where $\mathrm{Sim}(p,p') \neq \mathrm{Sim}(p',p)$.
	\begin{equation}
		\label{eq:sim_matrix}
		\mathrm{Sim}(p,p') =
		\sqrt{
			\frac{1}{T} \sum_{j \in \mathcal{T}} S^{p \rightarrow p'}_{j}
		}
	\end{equation}
	
	\subsubsection{Adaptive-$k$ Retrieval} We leverage the structure of $\mathrm{Sim}(p,p')$ to develop an adaptive-$k$ retrieval method that adapts to the query. The self-similarity $s_p = \mathrm{Sim}(p,p)$ of a page is used to rank the pages and define a candidate relevant set $\mathcal{R}_k$, containing the $k$ highest-scoring pages. The remaining $N-k$ pages constitute the irrelevant set $\mathcal{I}_k$. For each page $p \in \mathcal{R}_k$, we compute its coherence with the relevant set, $c^{p}_{k}$, and the leakage from the irrelevant set, $l^{p}_{k}$, as the mean similarity to pages in $\mathcal{R}_k$ and $\mathcal{I}_k$, respectively (\autoref{eq:p_in_r_i}). The scores $s_p$ are normalized to yield $w_{s_p}$ summing to one, and combined with $c^{p}_k$ and $l^{p}_k$ in the cost function $\mathcal{J}(k)$. In addition, ViSAR is budget-aware and takes as input the maximum number of pages Max-$k$ that can be passed to the LVLM. Therefore, $k^\star$ is obtained by minimizing $\mathcal{J}(k)$ over $1 \leq k \leq \min(\text{Max-}k,N)$ (\autoref{eq:k_star}). This requires the evaluation of at most $\min(\text{Max-}k,N)$ candidate sets, as pages are considered in order of self-similarity, while an exhaustive search would require evaluating $2^N$ sets. Consequently, the budget is only maximised when the unconstrained optimum of $\mathcal{J}(k)$ equals or exceeds Max-$k$, in which case the selected $k^\star$ is limited to the available budget.
	
	\begin{equation}
		\label{eq:p_in_r_i}
		c^{p}_{k} = \mathrm{mean}_{p' \in \mathcal{R}_k} \big[ \mathrm{Sim}(p,p') \big]
	\end{equation}
	
	\begin{equation}
		l^{p}_{k} = \mathrm{mean}_{p' \in \mathcal{I}_k} \big[ \mathrm{Sim}(p,p') \big]
	\end{equation}
	
	\begin{equation}
		\label{eq:j_k}
		\mathcal{J}(k) = \sum_{p \in \mathcal{R}_k} w_{s_p} \Big( c^{p}_{k} - \gamma \, l^{p}_{k} \Big)
	\end{equation}
	
	\begin{equation}
		\label{eq:k_star}
		k^\star = \operatorname*{arg\,min}_{1 \leq k \leq \min(\text{Max-}k,N)} \mathcal{J}(k)
	\end{equation}
	
	The parameter $\gamma > 0$ controls the leakage penalty. Also, since the minimum $\mathcal{J}(k^\star)$ may imply residual leakage from pages in $\mathcal{I}_{k^\star}$, we evaluate whether $k^\star$ corresponds to a sharp transition by accepting $\mathcal{R}_{k^\star + 1}$ only if $k^\star + 1 \leq \text{Max-}k$ and $\mathcal{J}$ varies more sharply beyond $k^\star$ than before it, indicating a large drop in leakage caused by the transitioning page (Algorithm~S1).
	
	\subsubsection{Implementation Details.} Since $\mathcal{J}(k)$ only involves the $\min(\text{Max-}k,N)$ pages with the highest self-similarity $s_p$, the computation of similarity matrix can be restricted to the corresponding entries $\mathrm{Sim}(p,\cdot)$. This requires first computing the diagonal entries $\mathrm{Sim}(p,p)$ to obtain the scores $s_p$, reducing the total number of page similarities to compute from $N^2$ to $(\text{Max-}k+1)N$. Additionally, we exploit the natural sparsity induced by ViSAR, since \autoref{eq:w_pj} produces a subset of inactive pages whose patch weights are all zero. These pages cannot contribute to the similarity matrix in \autoref{eq:sim_matrix} and are excluded from the computation of \autoref{eq:sim_ppj} and the optimization of $\mathcal{J}(k)$. Consequently, it implies that the number $\min(\text{Max-}k,N)$ of sets evaluated by $\mathcal{J}(k)$ and the number $(\text{Max-}k+1)N$ of similarities computed in the similarity matrix, are upper bounds as inactive pages are ignored. Finally, \autoref{eq:sim_ppj} is evaluated block-wise rather than as a dense tensor, reducing peak memory usage. These implementation choices preserve \emph{exact} mathematical equivalence while avoiding unnecessary computation.
	
	\section{Experiments}
	
	We evaluate four research questions: (RQ1) Can ViSAR effectively adapt the number of retrieved pages while preserving answer accuracy, compared with fixed top-$k$ retrieval and other adaptive-$k$ retrieval methods? (RQ2) How does ViSAR impact retrieval and generation latencies across budgets and documents? (RQ3) How does ViSAR generalize across configurations? (RQ4) Can ViSAR's similarity matrix provide insights into answer quality?
	
	\subsection{Experimental Setup}
	
	All experiments were conducted on an NVIDIA A6000 GPU (48 GB of memory). We set $T=50$ in \autoref{eq:sim_matrix} and $\gamma = 10^{5}$ in \autoref{eq:j_k}. Additional details on the compute resources are provided in the \hyperref[supplementals]{supplementary material}.
	
	\textbf{Datasets.} MMLongBench \cite{MMLongBench} and LongDocURL \cite{LongDocURL} datasets were used, which provide the answer evidence pages for page ranking evaluation. They cover multiple scenarios to evaluate the answer generation, requiring textual and visual reasoning across pages, and efficient retrieval over multi-page documents.
	
	\textbf{Encoders.} We evaluate ViSAR on multi-vector embeddings from three OCR-free visual encoders: ColPali \cite{Colpali}, ColQwen2.5 \cite{Colpali}, and ColModernVBERT \cite{ModernVBert}. For broader comparison, we consider ColBERTv2 \cite{colbertv2}, a multi-vector text encoder operating on OCR-extracted text, and VisRAG-Ret \cite{visrag}, a single-vector visual encoder.
	
	\textbf{Retrieval Baselines.} Multi-vector embeddings use late-interaction with fixed top-$k$, whereas single-vector embeddings use cosine similarity with fixed top-$k$ \cite{visrag}. In addition, we consider two adaptive-$k$ heuristics originally proposed for late-interaction text retrieval: \emph{Largest-Gap} \cite{AdaptiveK} and \emph{Score-Cluster} \cite{CulsterAdaptiveK}. Both operate on the sorted one-dimensional list of page scores obtained via late-interaction, and are therefore directly transferable to visual late-interaction encoders, which follow the same scoring mechanism. Largest-Gap places the cutoff at the largest difference between consecutive ranked scores, i.e., $k^\star=\operatorname*{arg\,max}_i\,\big(\operatorname{score}(i)-\operatorname{score}(i+1)\big)$, where $\operatorname{score}(i)$ is the $i$-th highest page score. Score-Cluster clusters the sorted scores with HDBSCAN, chosen in preliminary experiments among the clustering algorithms of the original method, with its hyperparameters selected automatically using the silhouette score. It then places the cutoff at the largest gap between consecutive clusters, defined as the difference between the lowest score of a cluster and the highest score of the next one.
	
	\textbf{Oracle.} Using the late-interaction ranking, the Oracle adapts the number of retrieved pages by selecting the smallest top-$k$ that contains all evidence pages. It provides an upper bound on adaptive-$k$ under a standard and fixed ranking, which may include irrelevant pages.
	
	\textbf{ViSAR Evaluation.} Unless stated otherwise, ViSAR, Largest-Gap, and Score-Cluster are run under a budget of Max-$k$ pages, the maximum number of pages passed to the LVLM. Fixed top-$k$ retrieval uses $\text{top-}k=\text{Max-}k$. We first evaluate the accuracy--cost trade-off of ViSAR, using Qwen2.5-VL-7B-Instruct (greedy decoding) as the primary LVLM, compared with fixed top-$k$, Largest-Gap, and Score-Cluster retrieval. We then analyze the adaptive retrieval behaviour of ViSAR without budget (i.e., $\text{Max-}k=N$), in comparison to Oracle, Largest-Gap, and Score-Cluster, reporting precision, recall, and F1-score at rank $k^\star$, as well as ViSAR's ranking quality against late-interaction ranking, using recall and normalized discounted cumulative gain (NDCG) at ranks 5 and 10. Next, we evaluate the retrieval, generation, and end-to-end (retrieval + generation) latencies of ViSAR and its dependence on the budget and on document length, followed by a comparison with fixed top-$k$, Largest-Gap, and Score-Cluster. We also examine the robustness of ViSAR with four additional LVLMs combined with each encoder, and its sensitivity to the hyperparameters $T$ and $\gamma$. Finally, we provide an ablation study to assess the contribution of each component of ViSAR.
	
	Answer accuracy is measured with LLM-as-a-judge evaluation \cite{LLMAsAJudgeSurvey,LLMASAJudgeEval}, using Qwen2.5-14B-Instruct \cite{Qwen25} as the evaluator. It receives the query, the reference answer, and the generated answer, and predicts a binary correctness based on semantic equivalence, using few-shot examples and constrained structured output. Differences with fixed top-$k$ are assessed with paired bootstrap 95\% confidence intervals over queries.
	
	\subsection{Experimental Results}
	
	Preliminary experiments revealed numerical instabilities using ColModernVBERT on a few LongDocURL pages (7.31\% of queries). Careful investigation confirmed that the issue occurred during encoding, independently of ViSAR. These samples were excluded from evaluation.
	
	\subsubsection{Accuracy vs Context Budget} 
	
	\begin{figure}[t!]
		\centering
		
		\includegraphics[width=\linewidth]{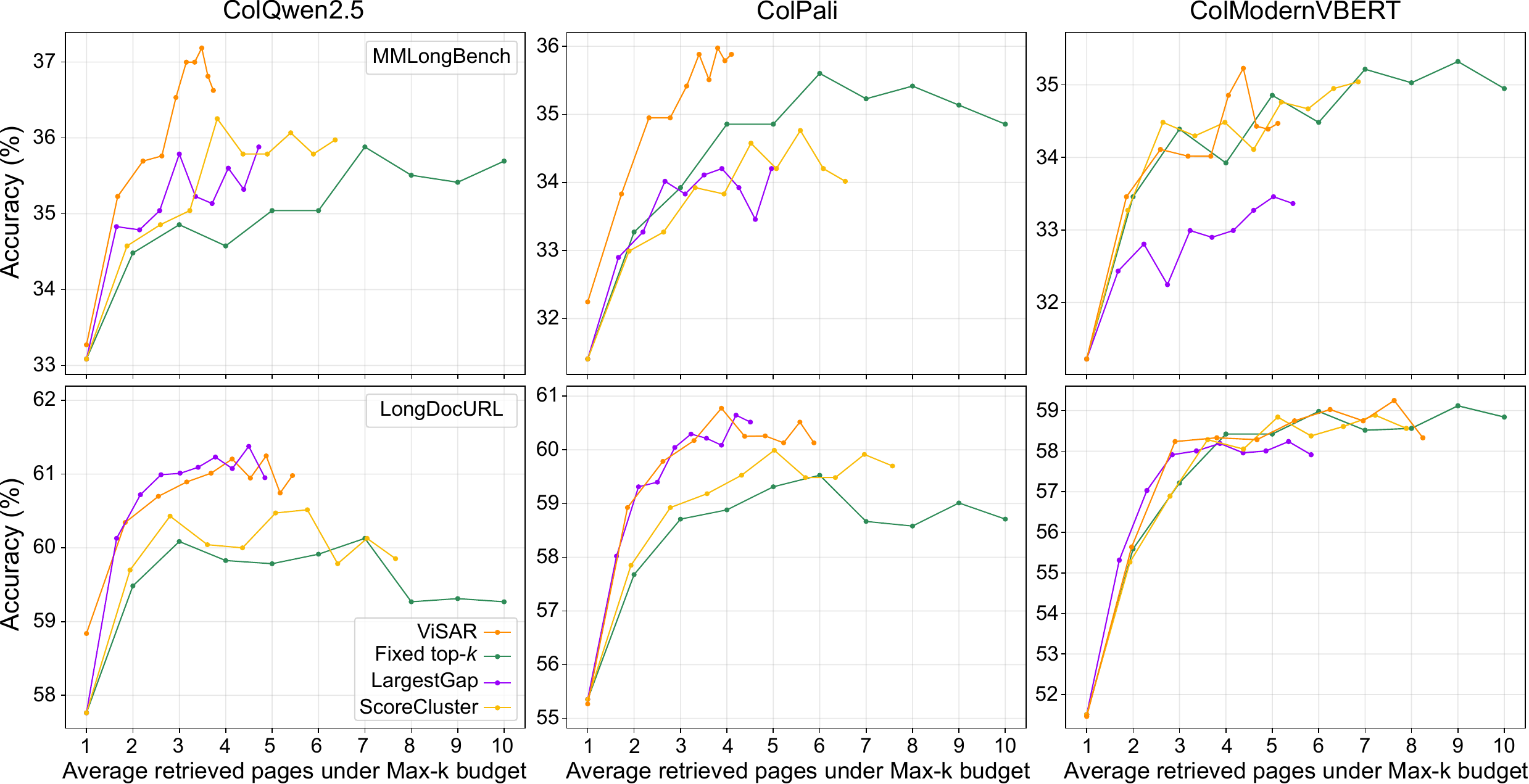}
		\vspace{-10pt}
		\caption{Accuracy--cost trade-off using Qwen2.5-VL-7B-Instruct. It shows the accuracy as a function of the average number of pages retrieved under Max-$k$ budgets from 1 to 10. Fixed top-$k$ retrieval always uses the full budget, while adaptive-$k$ retrievals only do so when required. Higher and further left is better. All adaptive methods reduce the number of pages compared to fixed top-$k$, but Largest-Gap and Score-Cluster do not reliably preserve accuracy, as fewer pages may come with lower accuracy in some configurations. In contrast, ViSAR reaches or exceeds fixed top-$k$ accuracy (values in \autoref{tab:accuracy_vs_numpages_all}). \label{fig:accuracy_vs_numpages_colqwen25_qwen25}}
		
		\vspace{1em}
		
		\small
		\tablecaption{Accuracy using Qwen2.5-VL-7B-Instruct under Max-5 and Max-10 budgets. ColBERTv2 (OCR-extracted text with Tesseract \cite{tesseract}) uses Qwen2.5-7B-Instruct. \# Pages denotes the average number of retrieved pages. ViSAR reaches or exceeds fixed top-$k$ accuracy with ColQwen2.5 and ColPali, and only yields marginal reductions with ColModernVBERT. Largest-Gap and Score-Cluster show less reliability across configurations. The number of pages retrieves depends on the configuration. ColQwen2.5 tends to retrieve the least number of pages, followed by ColPali and ColModernVBERT. \label{tab:accuracy_vs_numpages_all}}
		\vspace{0.5em}
		\setlength{\tabcolsep}{0.8mm}
		\renewcommand{\arraystretch}{0.925}
		\begin{tabularx}{\textwidth}{
				>{\raggedright\arraybackslash}p{0.14\textwidth}
				>{\raggedright\arraybackslash}p{0.14\textwidth}
				*{8}{>{\centering\arraybackslash}X}
			}
			\toprule
			
			& & \multicolumn{4}{c}{\textbf{MMLongBench}} & \multicolumn{4}{c}{\textbf{LongDocURL}} \\
			
			\cmidrule(lr){3-6} \cmidrule(lr){7-10}
			
			\textbf{Encoder} & \textbf{Retrieval} & Max-5 & \# Pages & Max-10 & \# Pages & Max-5 & \# Pages & Max-10 & \# Pages \\
			\midrule
			
			ColModern & Fixed top-$k$         & 34.75 & 5.0 & 34.95 & 10.0 & 58.42 & 5.0 & 58.84 & 10.0 \\
			VBERT     & Largest-Gap           & 32.99 & 3.2 & 33.36 & 5.4  & 58.00 & 3.4 & 57.91 & 5.8 \\
			& Score-Cluster         & 34.48 & 4.0 & 35.04 & 6.9  & 58.05 & 4.4 & 58.56 & 7.9 \\		
			& ViSAR (\textbf{ours}) & 34.01 & 3.7 & 34.48 & 5.1  & 58.28 & 4.6 & 58.33 & 8.2 \\
			
			\midrule
			
			ColPali & Fixed top-$k$         & 34.86 & 5.0 & 35.08 & 10.0 & 59.31 & 5.0 & 58.71 & 10.0 \\		
			& Largest-Gap           & 33.83 & 3.1 & 34.20 & 5.0  & 60.04 & 2.9 & 60.47 & 4.5 \\
			& Score-Cluster         & 33.83 & 3.9 & 34.02 & 6.6  & 59.53 & 4.3 & 59.70 & 7.6 \\			
			& ViSAR (\textbf{ours}) & 35.42 & 3.1 & 35.88 & 4.1  & 60.77 & 3.9 & 60.13 & 5.8 \\		
			
			\midrule
			
			ColQwen2.5 & Fixed top-$k$         & 35.04 & 5.0 & 35.69 & 10.0 & 59.79 & 5.0 & 59.27 & 10.0 \\
			& Largest-Gap           & 35.79 & 3.0 & 35.88 & 4.7  & 61.01 & 3.0 & 60.89 & 4.8 \\
			& Score-Cluster         & 36.25 & 3.8 & 35.97 & 6.4  & 60.00 & 4.4 & 59.83 & 7.7 \\		
			& ViSAR (\textbf{ours}) & 36.53 & 2.9 & 36.63 & 3.7  & 61.06 & 3.7 & 60.97 & 5.4 \\		
			
			\midrule
			
			ColBERTv2  & Fixed top-$k$ & 24.51 & 5.0 & 24.70 & 10.0 & 47.18 & 5.0 & 47.70 & 10.0 \\
			M3DocRAG   & Fixed top-$k$ & 34.86 & 5.0 & 35.08 & 10.0 & 59.31 & 5.0 & 58.71 & 10.0 \\
			VisRAG-Ret & Fixed top-$k$ & 34.48 & 5.0 & 35.69 & 10.0 & 57.29 & 5.0 & 58.02 & 10.0 \\
			
			\bottomrule
		\end{tabularx}
	\end{figure}

	\autoref{fig:accuracy_vs_numpages_colqwen25_qwen25} compares adaptive-$k$ retrieval with fixed top-$k$ retrieval methods using Qwen2.5-VL-7B-Instruct LVLM. It shows the average number of pages required to reach a given accuracy under a Max-$k$ budget, where fixed top-$k$ always uses the full budget and adaptive-$k$ only does so when required. ViSAR, Largest-Gap, and Score-Cluster all reduce the average number of retrieved pages compared to fixed top-$k$ while globally preserving the accuracy, but with varying reliability. Largest-Gap and Score-Cluster do not preserve accuracy in some configurations (e.g. ColPali and ColModernVBERT on MMLongBench), as fewer pages may come with lower accuracy. In contrast, ViSAR matches and often exceeds fixed top-$k$ accuracy with ColQwen2.5 and ColPali across all settings. With ColModernVBERT, although results are mixed, the trend across Max-$k$ budgets remains close to fixed top-$k$, with marginal to no accuracy costs relative to the substantial reductions in the average number of retrieved pages. Overall, these results indicate that adaptive-$k$ retrieval can efficiently reduce context size, with ViSAR providing more consistent results.
	
	\begin{figure}[t!]
		\centering
		\includegraphics[width=0.9\columnwidth]{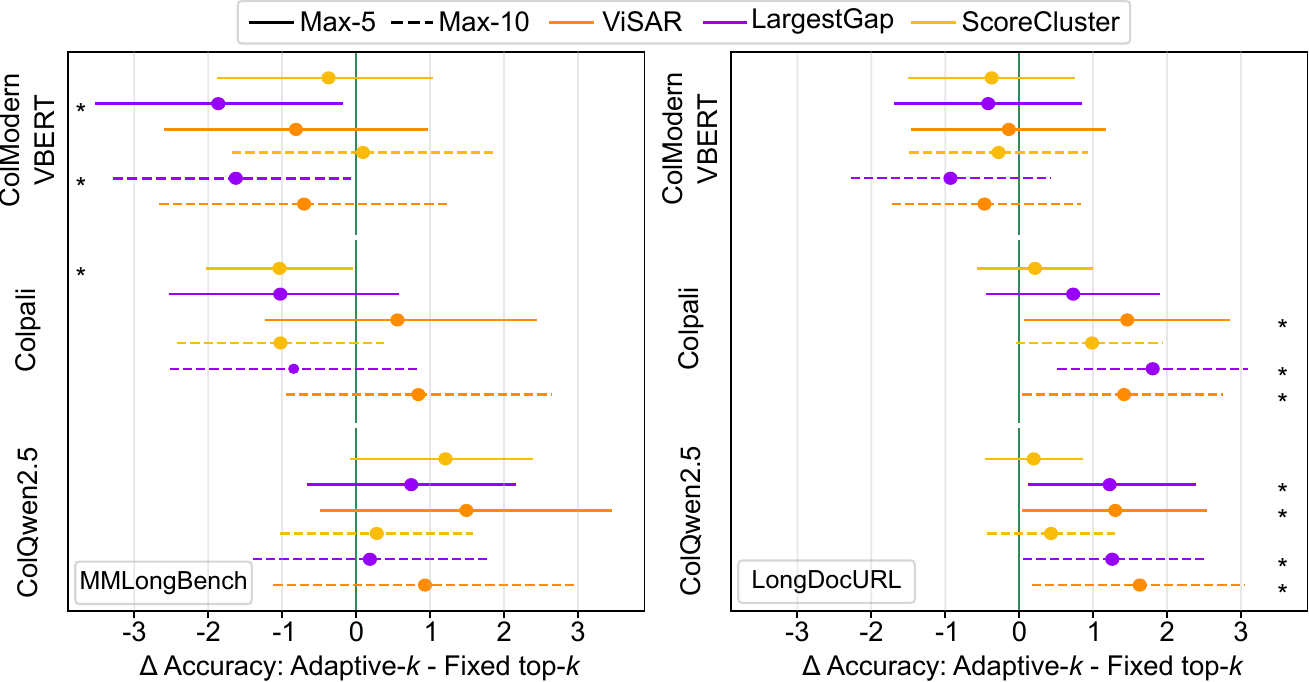}
		\caption{Accuracy differences between adaptive-$k$ and fixed top-$k$ using Qwen2.5-VL-7B-instruct under Max-5 and max-10 budgets. Mean differences are reported along with 95\% paired bootstrap confidence intervals (CI). Statistical significance is achieved when the CI does not contain zero, and is indicated by \textbf{*}. ViSAR exhibits no statistically significant loss and occasional significant improvement. Largest-Gap and Score-Cluster are less stable, with occasional significant losses.\label{fig:adaptive_k_and_fixed_k_diffs}}
		
	\end{figure}
	
	For precise values, \autoref{tab:accuracy_vs_numpages_all} reports the numbers for Max-5 and Max-10 budgets. Across encoders, datasets, and budgets, ViSAR matches or exceeds fixed top-$k$ using ColQwen2.5 and ColPali, is always among the best adaptive-$k$ methods, and falls only marginally below fixed top-$k$ using ColModernVBERT (at most 0.74 points). In contrast, Largest-Gap and Score-Cluster show more variability across configurations, either ranking among the best alongside ViSAR (e.g. Largest-Gap with ColQwen2.5 on LongDocURL) or among the worst (e.g. Score-Cluster with ColPali on MMLongBench). More specifically, \autoref{fig:adaptive_k_and_fixed_k_diffs} exhibits the difference between each adaptive-$k$ retrieval and fixed top-$k$ retrieval, with 95\% paired-bootstrap confidence intervals. Across 12 configurations (1 LVLM $\times$ 3 encoders $\times$ 2 budgets $\times$ 2 datasets), ViSAR yields a statistically significant improvement in 4 and shows no significant difference in the remaining 8, meaning that ViSAR never shows a significant loss with possible gain. In contrast, Largest-Gap and Score-Cluster occasionally show statistically significant losses.
	
	\autoref{tab:accuracy_vs_numpages_all} also reports accuracies for literature baselines ColBERTv2, M3DocRAG, and VisRAG-Ret (M3DocRAG uses ColPali with fixed top-$k$ and is included as a reference for state-of-the-art visual retrieval). ColBERTv2 shows a clear drop in accuracy compared to visual retrieval, consistent with the information lost in OCR-based pipelines. In contrast, ViSAR reaches the accuracy of existing literature baselines overall, while enabling a substantial reduction in the average number of retrieved pages and therefore input cost. Furthermore, the accuracy differences between ViSAR and fixed top-$k$ in \autoref{fig:adaptive_k_and_fixed_k_diffs}, show that ViSAR yields no statistically significant loss in accuracy and occasional significant improvement.
	
	\subsubsection{Adaptive Retrieval behaviour} 
	
	\begin{table}[h!]
		\centering
		
		\small
		\caption{Adaptive-$k$ retrieval behaviour under no budget (i.e, $\text{Max-}k=N$): mean ($\overline{m}$), median ($md$), Recall@$k^\star$ (R@$k^\star$), Precision@$k^\star$ (P@$k^\star$), and F1@$k^\star$. Oracle retrieves substantially more pages than the number of evidence pages, reflecting the known limitations \cite{ModernVBert,LateInteractionDefault} of late-interaction ranking. Adaptive methods balance recall and precision by adjusting the number of retrieved pages. ViSAR favours more compact retrieved sets on average, avoiding unnecessarily large sets. Full distributions are shown in \autoref{fig:distributions_k_star_ViSAR_Oracle}--\ref{fig:distributions_k_star_heuristic_methods}.\label{tab:retrieval_stats}}
		\setlength{\tabcolsep}{1.4mm}
		\renewcommand{\arraystretch}{0.95}
		\begin{tabularx}{\textwidth}{
				>{\raggedright\arraybackslash}p{0.11\textwidth}
				>{\raggedright\arraybackslash}p{0.14\textwidth}
				*{10}{c}
			}
			\toprule
			&
			& \multicolumn{5}{c}{\textbf{MMLongBench}} 
			& \multicolumn{5}{c}{\textbf{LongDocURL}} \\
			\cmidrule(lr){3-7}
			\cmidrule(lr){8-12}
			\textbf{Encoder} & \textbf{Retrieval}
			& $\overline{m}$ & $md$ & R@$k^\star$ & P@$k^\star$ & F1@$k^\star$
			& $\overline{m}$ & $md$ & R@$k^\star$ & P@$k^\star$ & F1@$k^\star$ \\
			
			\midrule
			ColQwen2.5 & Oracle                & 8.3  & 2 & 100.0 & 67.30 & 80.45 & 10.7 & 3  & 100.0 & 61.07 & 75.83 \\
			& Score-Cluster         & 18.6 & 8 & 85.57 & 29.51 & 43.89 & 36.6 & 20 & 89.49 & 18.85 & 31.14 \\
			& Largest-Gap           & 15.4 & 2 & 81.12 & 45.13 & 58.00 & 28.7 & 3  & 82.09 & 45.15 & 58.26 \\
			& ViSAR (\textbf{ours}) & 4.7  & 3 & 75.16 & 50.37 & 60.32 & 7.9  & 5  & 81.24 & 38.41 & 52.16 \\
			
			\midrule
			ColPali & Oracle                & 8.8  & 2 & 100.0 & 62.16 & 76.67 & 12.4 & 3  & 100.0 & 57.70 & 73.18 \\
			& Score-Cluster         & 18.7 & 8 & 82.64 & 26.12 & 39.70 & 35.8 & 18 & 87.56 & 19.15 & 31.43 \\
			& Largest-Gap           & 16.2 & 3 & 79.16 & 41.94 & 54.83 & 24.7 & 2  & 79.37 & 46.79 & 58.87 \\
			& ViSAR (\textbf{ours}) & 5.3  & 3 & 73.57 & 46.11 & 56.69 & 8.1  & 6  & 80.16 & 34.38 & 48.12 \\
			
			\midrule
			ColModern & Oracle                & 10.5 & 2  & 100.0 & 58.87 & 74.11 & 13.0 & 3  & 100.0 & 53.35 & 69.58 \\
			VBERT     & Score-Cluster         & 23.0 & 12 & 84.28 & 23.59 & 36.86 & 47.4 & 44 & 87.90 & 15.61 & 26.51 \\
			& Largest-Gap           & 20.0 & 4  & 80.57 & 35.81 & 49.58 & 39.3 & 4  & 83.57 & 34.94 & 49.28 \\
			& ViSAR (\textbf{ours}) & 7.5  & 4  & 74.07 & 37.43 & 49.73 & 13.5 & 11 & 83.99 & 17.47 & 28.92 \\
			
			\midrule
			\multicolumn{2}{l}{evidence pages}     & 1.9  & 1  & --    & -- & -- & 1.9 & 2 & -- & -- & -- \\
			
			\bottomrule
		\end{tabularx}

		\vspace{1em}
		
		\includegraphics[width=\linewidth]{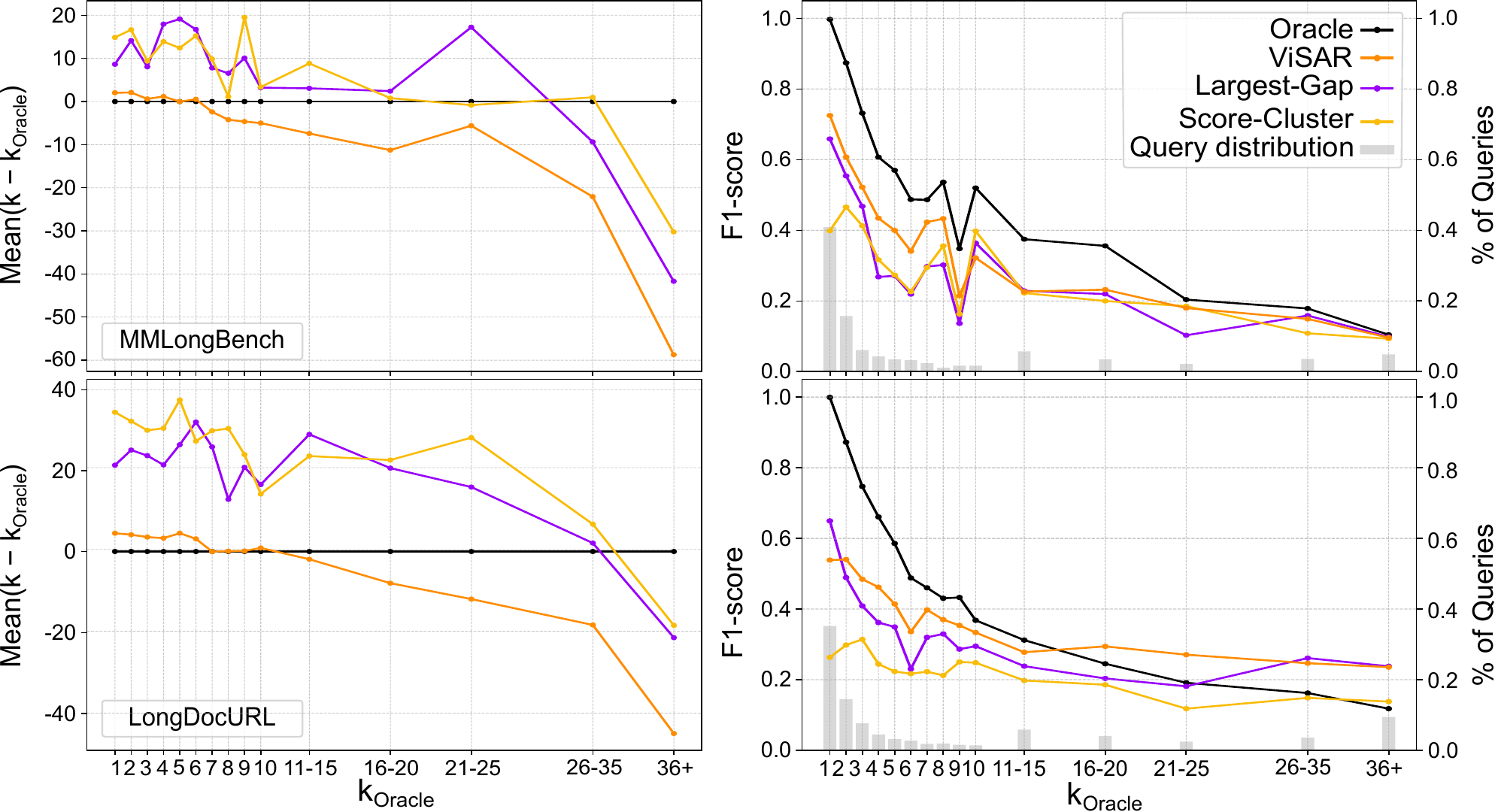} \\
		\figurecaption{Adaptive-$k$ retrieval, under no budget (i.e $\text{Max-}k=N$), as a function of $k_{\text{Oracle}}$ using ColQwen2.5 (Figures~S3--S4: consistent trends across encoders). \textbf{Left}: mean difference $k - k_{\text{Oracle}}$. ViSAR closely follows Oracle at low $k_{\text{Oracle}}$ and retrieves fewer pages as $k_{\text{Oracle}}$ increases, limiting the retrieval of irrelevant pages. Largest-Gap and Score-Cluster over-retrieve for most $k_{\text{Oracle}}$. \textbf{Right}: F1-score for each $k_{\text{Oracle}}$ with the percentage of queries. On MMLongBench, ViSAR achieves a higher F1, especially at lower $k_{\text{Oracle}}$. On LongDocURL, ViSAR outperforms Largest-Gap for most values, except $k_{\text{Oracle}}=1$.\label{fig:apdative_k_vs_k_oracle}}
	\end{table}

	We evaluate adaptive-$k$ retrieval under no budget constraint (i.e, $\text{Max-}k=N$), where \autoref{tab:retrieval_stats} reports mean, median, Recall@$k^\star$, Precision$@k^\star$, and F1$@k^\star$ across encoders for ViSAR, Oracle, Largest-Gap, and Score-Cluster. The mean number of pages retrieved by Oracle, compared to the mean number of evidence pages, shows that retrieving all evidence pages often requires significantly more pages than the number of evidence pages. In contrast, the lower median of Oracle indicates that compact retrieval is also possible. This behaviour reflects the known limitations of late-interaction ranking \cite{ModernVBert,LateInteractionDefault}, which may struggle to capture all relevant pages for complex queries, requiring the retrieval of intermediate irrelevant pages which can lead to unnecessarily large contexts. Adaptive retrieval methods exhibit a similar pattern to Oracle but explicitly balance Recall and Precision, with ViSAR retrieving fewer pages than Oracle on average while Largest-Gap and Score-Cluster retrieve substantially more.
	
	\autoref{fig:apdative_k_vs_k_oracle} left panels further characterise this behaviour as a function of $k_{\text{Oracle}}$. ViSAR closely follows Oracle at low $k_{\text{Oracle}}$ and retrieves fewer pages as $k_{\text{Oracle}}$ increases, limiting the inclusion of irrelevant pages that accumulate in complex queries. Largest-Gap and Score-Cluster over-retrieve across most $k_{\text{Oracle}}$, consistent with their higher mean $k^\star$ (\autoref{tab:retrieval_stats}). Consequently, ViSAR tends to favour precision, while Largest-Gap and Score-Cluster tend to favour recall at the expense of additional irrelevant pages.
	
	Dataset-level F1-scores (\autoref{tab:retrieval_stats}) also show that ViSAR performs best on MMLongBench, Largest-Gap performs best on LongDocURL, and Score-Cluster always performs worst. However, since the F1-score depends on the query distribution, \autoref{fig:apdative_k_vs_k_oracle} right panels report F1-score for each $k_{\text{Oracle}}$. It shows that ViSAR achieves higher F1-scores on MMLongBench, particularly for lower $k_{\text{Oracle}}$, which account for most queries. On LongDocURL, ViSAR outperforms Largest-Gap for most $k_{\text{Oracle}}$, except $k_{\text{Oracle}}=1$, suggesting that Largest-Gap is mostly effective for single-page queries, which constitute the majority of queries. This contributes to its higher dataset-level F1-score, whereas ViSAR provides a more consistent adaptive retrieval across queries. Overall (\autoref{fig:apdative_k_vs_k_oracle}, \autoref{fig:Metric_vs_k_oracle_F1-score}), ColQwen2.5 shows the best results, followed by ColPali, while ColModernVBERT exhibits a weaker trend. 
	
	\begin{table}[h!]
		\centering
		\caption{Page ranking evaluation, comparing ViSAR and late-interaction at ranks 5 and 10 on MMLongBench (similar trends for LongDocURL in \autoref{tab:ranking_results_both_datasets}). ViSAR consistently improves page ranking, providing stronger support for its adaptive-$k$ retrieval mechanism.\label{tab:ranking_results_MMLongBench}}
		\setlength{\tabcolsep}{1mm}
		\renewcommand{\arraystretch}{0.95}
		\begin{tabularx}{\columnwidth}{
				>{\raggedright\arraybackslash}p{0.34\columnwidth}
				*{4}{>{\centering\arraybackslash}X}
			}
			\toprule
			
			
			& \multicolumn{2}{c}{Recall} & \multicolumn{2}{c}{NDCG} \\
			\cmidrule(lr){2-3}
			\cmidrule(lr){4-5}	
			
			\textbf{Ranking Method} & @5 & @10 & @5 & @10 \\
			
			\midrule
			\multicolumn{5}{l}{\textit{Using Colpali}} \\
			\midrule
			
			\hspace{5pt} Late-Interaction 	   & 75.00          & 84.49          & 0.730          & 0.746          \\
			\hspace{5pt} ViSAR (\textbf{ours}) & \textbf{76.36} & \textbf{86.68} & \textbf{0.734} & \textbf{0.756} \\
			
			\midrule
			\multicolumn{5}{l}{\textit{Using ColQwen2.5}} \\
			\midrule
			
			\hspace{5pt} Late-Interaction      & 78.60          & 86.82          & 0.780          & 0.793          \\
			\hspace{5pt} ViSAR (\textbf{ours}) & \textbf{79.73} & \textbf{87.73} & \textbf{0.790} & \textbf{0.799} \\
			
			\midrule
			\multicolumn{5}{l}{\textit{Using ColModernVBERT}} \\
			\midrule
			
			\hspace{5pt} Late-Interaction      & 72.79          & 83.14          & 0.704          & 0.723          \\
			\hspace{5pt} ViSAR (\textbf{ours}) & \textbf{73.66} & \textbf{83.56} & \textbf{0.709} & \textbf{0.725} \\
			
			\bottomrule
		\end{tabularx}
	\end{table}
	
	\autoref{tab:ranking_results_MMLongBench} confirms this trend, showing that ViSAR's adaptive behaviour is grounded in improved ranking, outperforming late-interaction ranking across Recall and NDCG at ranks 5 and 10, indicating that its weighting mechanism effectively aligns page scoring with the query. Yet, ColModernVBERT shows modest improvements, consistent with its smaller size (250M vs. 3B parameters), which likely limits its ability to separate semantics, resulting in less compact sets and lower ranking performance overall (\autoref{tab:retrieval_stats}).
	
	\subsubsection{Latency and Overhead}
	
	\begin{figure}[h!]
		\centering
		
		\includegraphics[width=\linewidth]{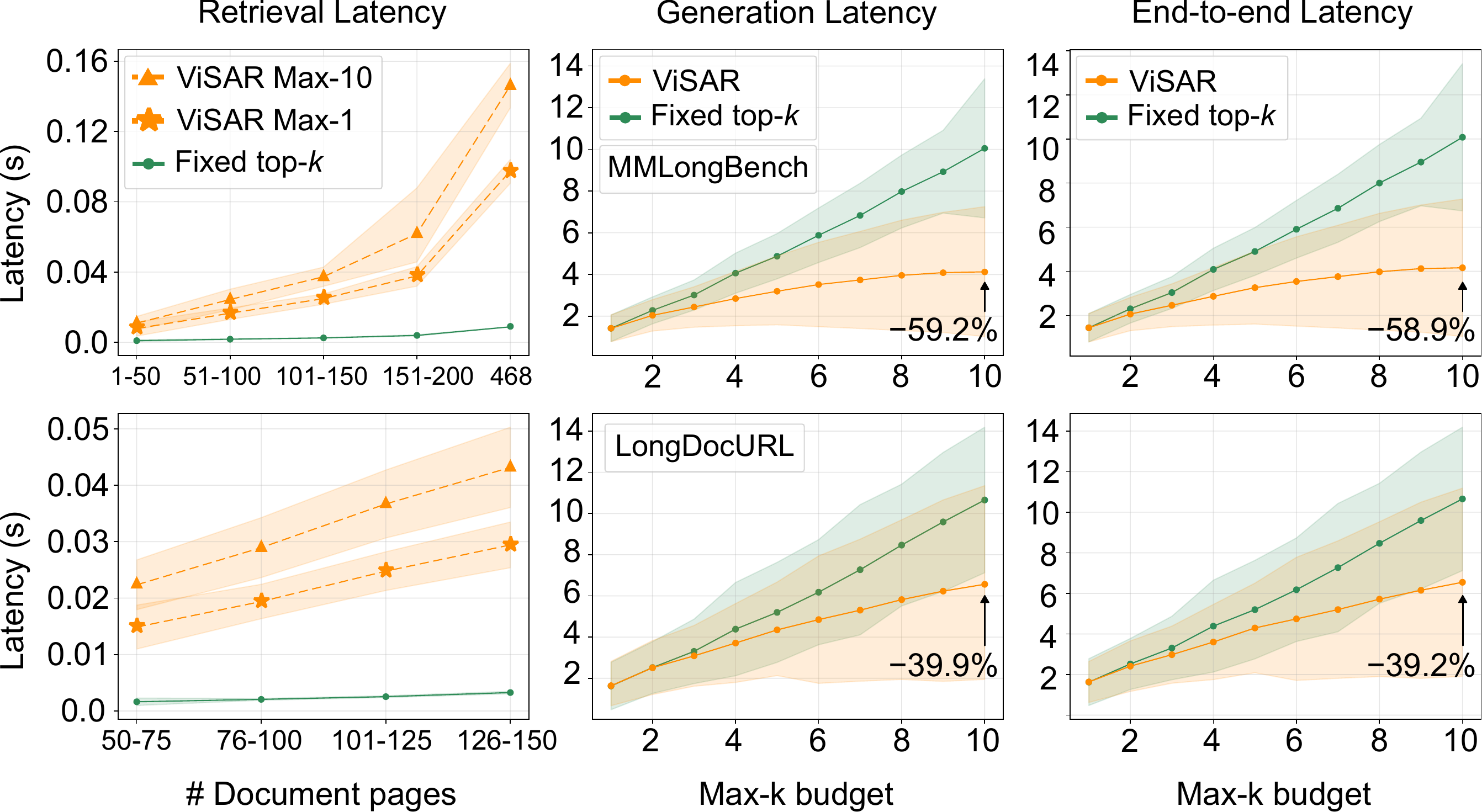}\\
		\caption{Average RAG latency in seconds under Max-$k$ budgets from 1 to 10, comparing ViSAR and fixed top-$k$ (shaded: std). ViSAR retrieval latency is shown for the lower and upper bounds Max-1 and Max-10. ViSAR introduces a retrieval overhead that grows with document size and budget but remains marginal, as ViSAR substantially reduces generation latency by limiting LVLM context. The net effect is an end-to-end latency reduction of up to 58.9\% for MMLongBench and 39.2\% for LongDocURL, at $k=10$.\label{fig:latency_RAG_VISAR}}

		\vspace{1em}
		
		\small
		\tablecaption{End-to-end (retrieval + generation) latency in seconds. Score-Cluster is consistently slower than ViSAR. Largest-Gap provides an advantage only on LongDocURL, which is expected to diminish at larger budgets given its substantially higher mean $k^\star$ (\autoref{tab:retrieval_stats}). A breakdown of retrieval and generation latencies is provided in \autoref{tab:latency_breakdown_visar_vs_heuristics}.\label{tab:latency_end-to-end_visar_vs_heuristics}}
		\setlength{\tabcolsep}{1mm}
		
		\begin{tabularx}{\columnwidth}{
				>{\raggedright\arraybackslash}p{0.28\columnwidth}
				*{4}{>{\centering\arraybackslash}X}
			}
			\toprule
			
			& \multicolumn{2}{c}{\textbf{MMLongBench}} & \multicolumn{2}{c}{\textbf{LongDocURL}} \\
			
			\cmidrule(lr){2-3} \cmidrule(lr){4-5}
			
			\textbf{Retrieval} & Max-5 & Max-10 & Max5 & Max-10 \\
			\midrule
			
			Fixed top-$k$         & 4.93 & 10.10 & 5.21 & 10.66 \\		
			Score-Cluster         & 4.60          & 6.73          & 4.86          & 8.36          \\
			Largest-Gap           & 3.35          & 5.45          & \textbf{3.76} & \textbf{5.63} \\
			ViSAR (\textbf{ours}) & \textbf{3.24} & \textbf{4.15} & 4.26          & 6.48          \\
			
			\bottomrule
		\end{tabularx}
	\end{figure}
	
	\begin{figure}[t!]
		\centering
		
		\includegraphics[width=\linewidth]{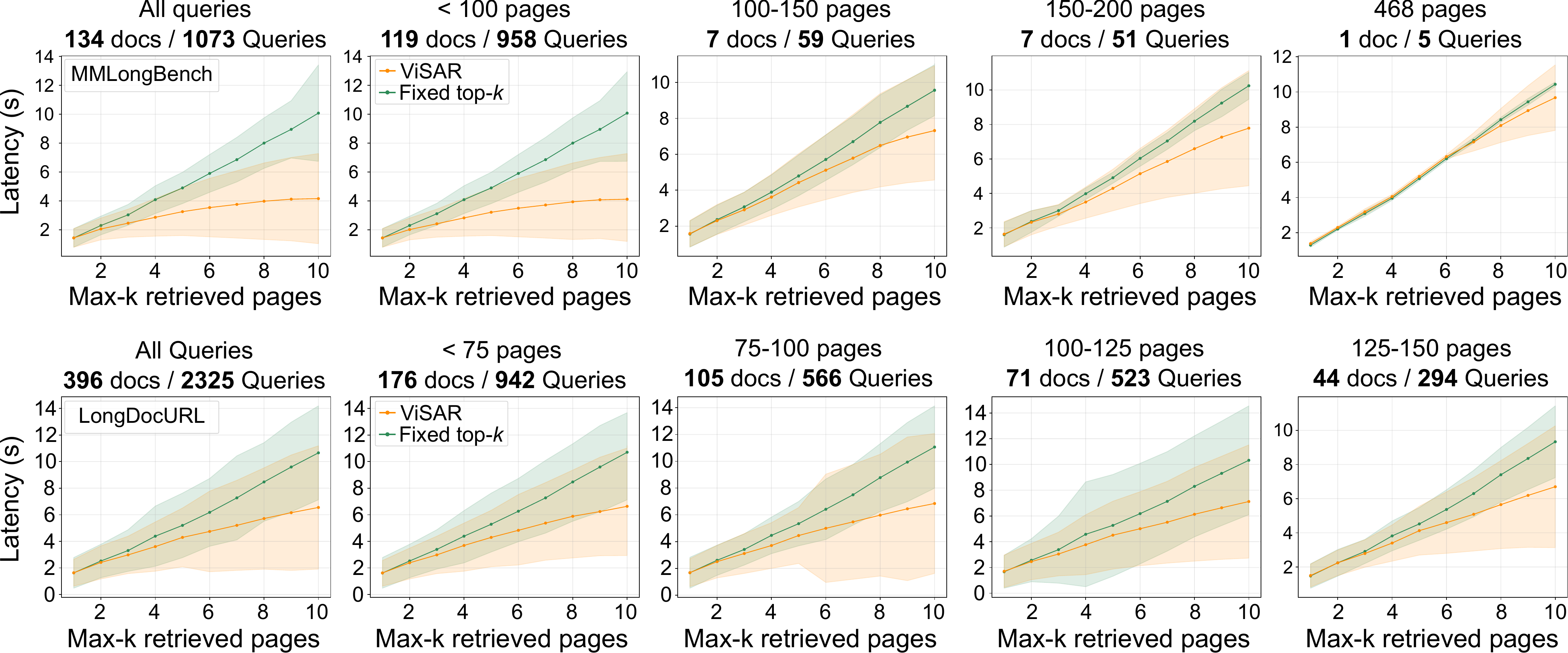}\\
		\caption{End-to-end (retrieval + generation) latency in seconds across document length, under Max-$k$ budgets from 1 to 10 (Shaded: std), and with the number of documents and queries reported for each group. ViSAR yields substantial latency gains for almost all documents length, with gains increasing with the budget, since a larger budget leaves more page to save. The latency gain decreases as document length increases, and only becomes negligible for the longest document (468 pages, MMLongBench), consistent with the average $k^\star$ that increases with document length (\autoref{tab:latency_end_to_end}).\label{fig:latency_RAG_ViSAR_higher_pages}}

		\vspace{1em}
		
		\small
		\tablecaption{Statistics of the selected number of pages $k^\star$ by document length, using ColQwen2.5 under no budget (i.e. $\text{Max-}k=N$). Docs/Queries reports the number of documents and queries in each group. The mean $k^\star$ increases with document length, while the average number of evidence pages ($\overline{\text{evidence pages}}$) remains low. The distribution of $k^\star$ also widens, but its 25th percentile and minimum remain small for most groups, showing that ViSAR still selects compact page sets for some queries on long documents, with the single 468 pages document representing the extreme case. This dependence on both document length and the query motivates further investigation of the similarity structure.\label{tab:latency_end_to_end}}
		\vspace{0.5em}
		\setlength{\tabcolsep}{0.7mm}
		\renewcommand{\arraystretch}{0.95}
		\begin{tabularx}{\linewidth}{
				>{\raggedright\arraybackslash}p{0.135\linewidth}
				*{4}{>{\centering\arraybackslash}X}			
				>{\centering\arraybackslash}p{0.05\linewidth}
				*{5}{>{\centering\arraybackslash}X}			
			}
			\toprule
			
			& \multicolumn{5}{c}{\textbf{MMLongBench}} & \multicolumn{5}{c}{\textbf{LongDocURL}} \\
			
			\cmidrule(lr){2-6} \cmidrule(lr){7-11}
			
			Pages        & All        & <100      & 100-150 & 150-200 & 468   & All        & <75     & 75-100    & 100-125  & 125-150  \\
			Docs/Queries & 134/1073 & 119/958 &  7/59 & 7/51  & 1/5 & 396/2325 & 176/942 & 105/566 & 71/523 & 44/294 \\
			
			\midrule
			
			mean            & 4.7 & 3.6 & 13.5 & 13.9 & 32.0 & 7.9  & 5.9 & 8.1  & 9.1  & 12.0 \\
			median          & 3.0 & 3.0 & 15.0 & 8.0  & 30.5 & 5.0  & 4.0 & 5.0  & 6.0  & 7.0  \\
			25th percentile & 1.0 & 1.0 & 5.0  & 3.0  & 22.0 & 2.0  & 2.0 & 2.0  & 3.0  & 3.0  \\
			75th percentile & 5.0 & 4.5 & 20.0 & 21.6 & 40.5 & 10.0 & 8.0 & 10.0 & 11.0 & 15.0 \\
			minimum         & 1   & 1   & 1    & 1    & 7    & 1    & 1   & 1    & 1    & 1    \\
			
			\midrule
			$\overline{\text{evidence pages}}$ & 1.9 & 1.9 & 1.6 & 1.9 & 1 & 1.9 & 1.8 & 1.9 & 1.8 & 2.1\\
			
			\bottomrule
		\end{tabularx}
		
	\end{figure}
	
	\textbf{Latency Reduction and Retrieval Overhead.} The reduction in context size enabled by adaptive retrieval lowers the LVLM processing cost. While fixed top-$k$ retrieval always saturates the budget, ViSAR dynamically adjusts the number of pages and only uses the full budget when required. \autoref{fig:latency_RAG_VISAR} reports retrieval, generation, and end-to-end (retrieval + generation) latencies, showing that ViSAR introduces a retrieval overhead that contributes only marginally, achieving a substantial reduction in generation latency that dominates the total cost. This results in an average end-to-end latency reduction of up to 58.9\% on MMLongBench and 39.2\% on LongDocURL at Max-10 budget. More specifically, as $\mathcal{J}(k)$ involves $\min(\text{Max-}k,N)$ pages, the computational cost of ViSAR increases both with the budget and the number of pages. However, \autoref{tab:latency_breakdown_visar_vs_heuristics} shows average retrieval latencies of 0.018 seconds and 0.030 seconds at Max-10 budget for MMLongBench and LongDocURL, respectively. Even for queries on the longest document evaluated (MMLongBench, 468 pages) at Max-10 budget, the retrieval overhead remains marginal at 0.15 seconds compared with generation latency.
	
	\textbf{Effect of Document Length.} \autoref{fig:latency_RAG_ViSAR_higher_pages} breaks down the end-to-end latency by document length. Almost all documents length benefit from a substantial latency reduction, with the gain that increases with the budget, since a larger budget leaves more page processing to save. The only exception is the longest document (468 pages, MMLongBench), which yields no latency gain at lower budgets and only small gains at higher budgets, despite marginal retrieval overheads. More specifically, the latency gain decreases as document length increases, and become negligible only for this document. This behaviour is consistent with the average $k^\star$ that increases with document length, whereas the average number of evidence pages stays low (\autoref{tab:latency_end_to_end}). A plausible explanation lies in the leakage term of $\mathcal{J}(k)$, which averages similarities over the $N-k$ pages of $\mathcal{I}_k$. In long documents, this mean is diluted: moving a page from $\mathcal{I}_k$ to $\mathcal{R}_k$ reduces the leakage by an amount proportional to $1/(N-k)$, causing the effect of a transitioning relevant page to shrink as $N$ increases. Consequently, more pages may need to transition into $\mathcal{R}_k$ for the leakage to drop significantly. However, in preliminary experiments, we tested several variants of the leakage term, including a weighted mean using the scores $s_p$ to reduce the influence of low self-similarity pages, and we obtained no clear improvement. It suggests that a more sophisticated mechanism is required for the leakage, which is further supported by \autoref{tab:latency_end_to_end}: although the average $k^\star$ increases with document length, its 25th percentile and minimum values remain small for a substantial amount of queries on long documents, where ViSAR still selects compact page sets that fit within the budget. Therefore, document length likely explains only partly the behaviour of the average $k^\star$.
	
	\textbf{Comparison with Adaptive-$k$ Baselines.} \autoref{tab:latency_end-to-end_visar_vs_heuristics} further compares the end-to-end latency of ViSAR with Largest-Gap and Score-Cluster. Score-Cluster remains consistently slower than ViSAR, whereas Largest-Gap provides an advantage only on LongDocURL, with gains of up to 47.2\% over fixed top-$k$ at Max-10 budget. However, this latency advantage only holds under the specific budgets evaluated. \autoref{fig:latency_RAG_VISAR} shows that fixed top-$k$ latency increases linearly with the number of pages sent to the LVLM, which means that the average latency of adaptive-$k$ methods follows the mean of $\min(k^\star , \text{Max-}k)$ over queries. Additionally, \autoref{tab:retrieval_stats} shows that Largest-Gap $k^\star$ distributions have a lower or comparable median than ViSAR but that Largest-Gap has a substantially higher mean, reflecting a heavy-tailed distribution that is truncated by the Max-10 budget. These two observations imply that Largest-Gap will likely become slower than ViSAR once the budget exceeds a threshold. However, this threshold lies beyond the range of budgets evaluated.
	
	\subsubsection{Robustness}
	
	\begin{table}[h!]
		\centering
		
		\small
		\caption{Accuracy (\%) comparing ViSAR and Fixed top-$k$ across four additional LVLMs: Qwen3-VL \cite{Qwen3vl}, InternVL3.5 \cite{InternVL}, Idefics3 \cite{Idefics3}, and LLaVA-OV \cite{LlavaOV}. It compares ViSAR with fixed top-$k$ retrieval at Max-5 and Max-10 budgets. Bold indicates statistically significant differences (95\% confidence interval in \autoref{fig:all_adaptive_k_and_fixed_diffs}). Across the 48 configurations, ViSAR yields a significant improvement in 23, with no significant difference in the remaining 25, and never shows a significant loss.\label{tab:answer_accuracy_additional_LVLMs}}	
		\setlength{\tabcolsep}{0.5mm}
		\begin{tabularx}{\textwidth}{
				>{\raggedright\arraybackslash}p{0.14\textwidth}
				>{\raggedright\arraybackslash}p{0.12\textwidth}
				>{\raggedright\arraybackslash}p{0.13\textwidth}
				*{11}{>{\centering\arraybackslash}X}
			}
			\toprule
			
			& &
			& \multicolumn{2}{c}{\begin{tabular}{c}Qwen3-VL\\8B-Instruct\end{tabular}}
			& \multicolumn{2}{c}{\begin{tabular}{c}InternVL3.5\\8B-Instruct\end{tabular}} 
			& \multicolumn{2}{c}{\begin{tabular}{c}Idefics3\\8B-Llama3\end{tabular}}
			& \multicolumn{2}{c}{\begin{tabular}{c}LLaVA-OV\\Qwen2-7B\end{tabular}} \\
			\cmidrule(lr){4-5}
			\cmidrule(lr){6-7}
			\cmidrule(lr){8-9}
			\cmidrule(lr){10-11}
			\textbf{Dataset} & \textbf {Encoder} & \textbf{Retrieval} 
			& Max-5 & Max-10 &Max-5 & Max-10 & Max-5 & Max-10 &Max-5 & Max-10 \\
			
			\midrule
			
			MMLongBench & ColModern & Fixed top-$k$ & 38.82 & 40.08 & 30.76 & 30.76 & 21.81 & 20.69 & 15.66 & 9.32 \\
			& VBERT & ViSAR (\textbf{ours}) & 37.94 & 39.79 & 30.57 & 31.31 & 20.93 & 20.88 & \textbf{18.27} & \textbf{16.59} \\
			
			\cmidrule(lr){2-11}
			
			& ColPali & Fixed top-$k$ & 38.65 & 39.51 & 32.15 & 30.76 & 21.76 & 19.66 & 15.94 & 9.97  \\
			& & ViSAR (\textbf{ours}) & 37.99 & 38.68 & 32.71 & 32.53 & 22.83 & \textbf{22.83} & \textbf{19.76} & \textbf{20.04} \\
			
			\cmidrule(lr){2-11}
			
			& ColQwen2.5 & Fixed top-$k$ & 40.08 & 40.73 & 31.69 & 29.64 & 22.27 & 20.60 & 16.03 & 10.25 \\
			& & ViSAR (\textbf{ours}) & 39.33 & 39.54 & \textbf{33.74} & \textbf{33.74} & \textbf{24.28} & \textbf{23.39} & \textbf{20.69} & \textbf{19.94} \\
			
			\midrule
			
			LongDocURL & ColModern & Fixed top-$k$ & 63.57 & 65.92 & 57.77 & 55.73 & 40.84 & 37.87 & 25.66 & 16.15 \\
			& VBERT & ViSAR (\textbf{ours}) & 63.02 & 66.07 & 58.10 & 56.94 & 41.02 & 38.52 & \textbf{26.98} & \textbf{20.97} \\
			
			\cmidrule(lr){2-11}
			
			& ColPali & Fixed top-$k$ & 64.56 & 67.36 & 58.68 & 55.74 & 40.77 & 37.76 & 26.41 & 17.03 \\
			& & ViSAR (\textbf{ours}) & 65.64 & 66.88 & 59.57 & \textbf{59.01} & 41.76 &\textbf{41.20} & \textbf{30.80} & \textbf{27.14}  \\		           
			
			\cmidrule(lr){2-11}
			
			& ColQwen2.5 & Fixed top-$k$ & 66.75 & 67.69 & 58.54 & 55.40 & 40.90 & 37.94 & 26.11 & 16.95 \\
			& & ViSAR (\textbf{ours}) & 67.27 & 67.18 & \textbf{60.68} & \textbf{60.17} & \textbf{43.02} & \textbf{41.59} & \textbf{32.56} & \textbf{30.32} \\
			
			\bottomrule
		\end{tabularx}
		
		\vspace{1em}
		
		\includegraphics[width=\linewidth]{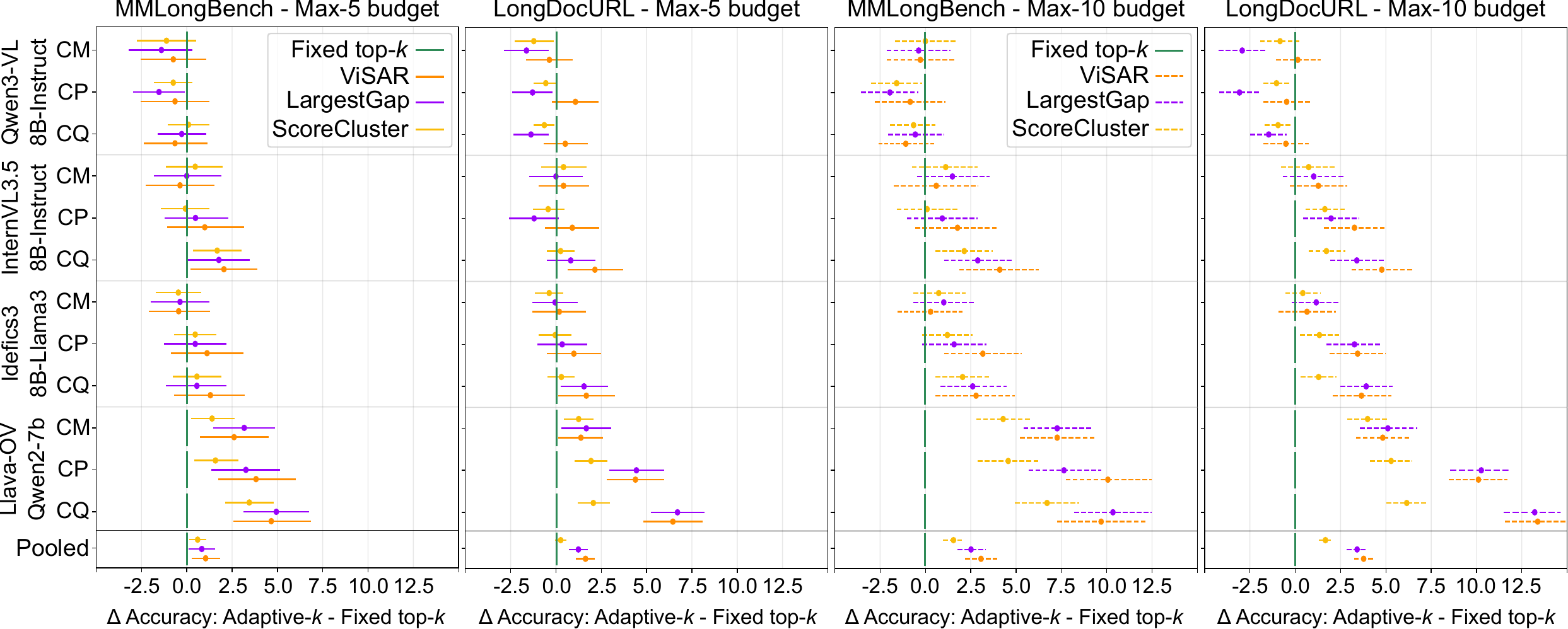}\\
		\figurecaption{Accuracy differences (\%) between adaptive-$k$ and fixed top-$k$ across four LVLMs: Qwen3-VL \cite{Qwen3vl}, InternVL3.5 \cite{InternVL}, Idefics3 \cite{Idefics3}, and LLava-OV \cite{LlavaOV}. LVLMs are evaluated using ColModernVBERT (CM), ColPali (CP), and ColQwen2.5 (CQ). The green lines marks fixed top-$k$. Points are mean differences with 95\% paired bootstrap confidence intervals (CI). A difference is statistically significant when its CI excludes zero. ``Pooled'' rows aggregate all LVLMs and encoders. ViSAR shows no significant loss and several significant gains, whereas Largest-Gap and Score-Cluster show both significant losses and gains. Pooled mean differences are higher for ViSAR than for Largest-Gap, although their CIs overlap. The largest gains occur for LLaVA-OV, the lowest-performing LVLM under fixed top-$k$ (\autoref{tab:answer_accuracy_additional_LVLMs}), which is more sensitive to irrelevant pages.\label{fig:all_adaptive_k_and_fixed_diffs}}	
		
	\end{table}
	
	\textbf{ViSAR generalization}. We assess ViSAR using four additional LVLMs, for each of the three encoders. \autoref{tab:answer_accuracy_additional_LVLMs} reports the accuracy of ViSAR and fixed top-$k$, and \autoref{fig:all_adaptive_k_and_fixed_diffs} shows the mean accuracy difference to fixed top-$k$ for ViSAR, Largest-Gap, and Score-Cluster, along with their 95\% paired bootstrap confidence intervals. Across 48 configurations (4 LVLMs $\times$ 3 encoders $\times$ 2 budgets $\times$ 2 datasets), ViSAR yields a statistically significant improvement in 23 and shows no significant difference in the remaining 25, meaning that ViSAR never shows a significant loss with possible gains (\autoref{tab:answer_accuracy_additional_LVLMs}). In contrast, Largest-Gap and Score-Cluster are less consistent, with both significant losses and significant gains (\autoref{fig:all_adaptive_k_and_fixed_diffs}). In addition, pooled over LVLMs and encoders, ViSAR obtains the highest mean difference in all four dataset--budget settings, substantially above Score-Cluster in three out of four settings, and modestly above Largest-Gap in the four settings but with overlapping confidence intervals.
	
	However, the gains are not uniform across LVLMs (\autoref{fig:all_adaptive_k_and_fixed_diffs}). The largest improvements are obtained with LLaVA-OV-Qwen2-7B, with up to $+13.37$ (\autoref{tab:answer_accuracy_additional_LVLMs}, LongDocURL, ColQwen2.5, Max-10). This LVLM shows a strong sensitivity to context size, as shown by its fixed top-$k$ accuracy that decreases when the budget increases (e.g LongDocURL, ColQwen2.5: from 26.11\% to 16.95\%), while ViSAR enables improving and stabilising accuracy (32.56\% to 30.32\%) by reducing the number of irrelevant pages. Similarly, InternVL3.5-8B-Instruct and Idefics3-8B-Llama3 show the same but milder trend. In contrast, Qwen3-VL-8B-Instruct benefits from a larger context, as shown by its fixed top-$k$ accuracy that increases when the budget increases (up to $+2.80$ points on LongDocURL with ColPali). For this LVLM, ViSAR shows no statistically significant changes relative to fixed top-$k$ accuracy, with differences ranging from $-1.19$ to $+1.08$ points, whereas Largest-Gap and Score-Cluster yield significant losses in accuracy on several configurations (e.g \autoref{fig:all_adaptive_k_and_fixed_diffs}, LongDocURL, ColPali, Max-10). Consequently, even though all adaptive-$k$ methods benefit LVLMs sensitive to irrelevant context, only ViSAR remains consistently close to fixed top-$k$ for LVLMs that better exploit long contexts, while substantially reducing both the number of retrieved pages (\autoref{tab:retrieval_stats}) and end-to-end RAG latency (\autoref{fig:latency_RAG_VISAR}).
	
	\textbf{Effect of encoders.} Across LVLMs, the three adaptive-$k$ methods tend to yield the best accuracy with ColQwen2.5, followed by ColPali and then ColModernVBERT (\autoref{fig:all_adaptive_k_and_fixed_diffs}). This ordering follows their ranking quality (\autoref{tab:ranking_results_MMLongBench}) and their average number of retrieved pages (\autoref{tab:retrieval_stats}). ColModernVBERT exhibits the weakest ranking quality and produces larger page sets on average, which may be due to its smaller size (250M parameters compared with 3B for ColPali and ColQwen2.5). Consequently, for all adaptive-$k$ methods, ColModernVBERT enables statistically significant accuracy gains only with LLaVA-OV-Qwen2-7B, the LVLM most sensitive to context size and irrelevant pages. Conversely, the trend across encoder is weaker for LVLMs that better exploit the input context, and is unclear when using Qwen3-VL-8B-Instruct, likely because of its ability to yield higher accuracies overall (\autoref{tab:answer_accuracy_additional_LVLMs}), making page set differences between encoders less critical. For all encoders, using Qwen3-VL-8B-Instruct also tends to yield slightly lower accuracies, but only ViSAR remains close to fixed top-$k$, while Largest-Gap and Score-Cluster yield statistically significant losses in some settings.
	
	\textbf{Hyperparameter sensitivity.} In all experiments, we used a single default $T=50$ and $\gamma = 10^{5}$ for every encoder, dataset, and budget.
	
	\autoref{fig:sensitivity_at_5_top_T_k_star} and \ref{fig:sensitivity_at_10_top_T_k_star} report the gain in ranking quality of ViSAR over late-interaction as a function of $T \in [1,10^3]$, the number of source-patch interactions used to compute the similarity matrix (\autoref{eq:sim_matrix}), for the two datasets and three encoders. ViSAR preserves or improves recall and NDCG at ranks 5 and 10 for $T\leq150$ in every configuration. Also, although the best $T$ depends on the encoder (around 10--25 for ColModernVBERT, 50--75 for ColQwen2.5, and above 100 for ColPali), the differences in ranking quality remain marginal. The default $T=50$ that we used costs at most 0.4 points of Recall while still yielding 1.37 gain (\autoref{tab:ranking_results_both_datasets}, \autoref{fig:sensitivity_at_5_top_T_k_star}, LongDocURL, ColPali, Rank 5: 75.07\% for Late-Interaction, 76.44\% (+1.37) for $\text{ViSAR}_{T=50}$, 76.84\% (+1.77) for $\text{ViSAR}_{T=300}$), and at most 0.007 NDCG while still yielding 0.004 gain (\autoref{tab:ranking_results_both_datasets}, \autoref{fig:sensitivity_at_5_top_T_k_star}, MMLongBench, ColPali, Rank 5: 0.730 for Late-Interaction, 0.734 (+0.004) for $\text{ViSAR}_{T=50}$, 0.741 (+0.011) for $\text{ViSAR}_{T=150}$). Overall, $T=50$ yields recall gains ranging from 0.75 to 2.19 points and NDCG gains from 0.004 to 0.013. Therefore, these results show that the query-conditioned weighting mechanism of ViSAR, designed to localize evidence and enable adaptive-$k$ retrieval, correctly aligns query and page embeddings, and is robust across encoders and datasets. Page ranking quality remains largely stable over a large range of $T$, with negligible variations across encoders and datasets, while enabling slight improvements.
	
	\autoref{fig:sensitivity_analysis_k_star_function_of_gamma} reports the behaviour of $k^\star$ as a function of $\gamma \in [1,10^8]$, the leakage penalty in the retrieval cost function $\mathcal{J}(k)$ (\autoref{eq:j_k}). It shows mean~$k^\star$ over queries and mean~$|\Delta k^\star|$ between consecutive values of $\gamma$, both under no budget (i.e $\text{Max-}k=N$). Small values of $\gamma$ lead to over-retrieval, with $\gamma=1$ yielding a mean~$k^\star$ across encoders of 13--18 pages on MMLongBench and 30--38 on LongDocURL. As $\gamma$ increases, $k^\star$ decreases and stabilises beyond $10^{4}$, around its current mean values (\autoref{tab:retrieval_stats}): for all encoders and datasets, the mean, median, and quartiles of $k^\star$ become flat and mean~$|\Delta k^\star|$ reaches close to zero for $\gamma\geq10^{4}$. Therefore, since $\mathcal{J}(k)$ considers pages in order of self-similarity $s_p$ independently of $\gamma$, mean~$|\Delta k^\star|$ close to zero implies little to no changes in retrieved pages and identical LVLM inputs within the plateau achieved with $\gamma \in [10^4,10^8]$. Mechanistically, $\gamma$ needs to be large enough for the leakage term to dominate the objective (\autoref{eq:j_k}), eventually reaching a point where $k^\star$ cannot decrease anymore: a smaller $k^\star$ would exclude relevant pages from the selected set, causing their leakage to outweigh the leakage of irrelevant pages and $\mathcal{J}(k)$ to further decrease, while a stable $k^\star$ implies no significant leakage anymore and $\mathcal{J}(k)$ to stop decreasing.
	
	\subsubsection{Similarity Matrix and Accuracy}
	
	\begin{figure}[h!]
		\centering
		\includegraphics[width=0.8\linewidth]{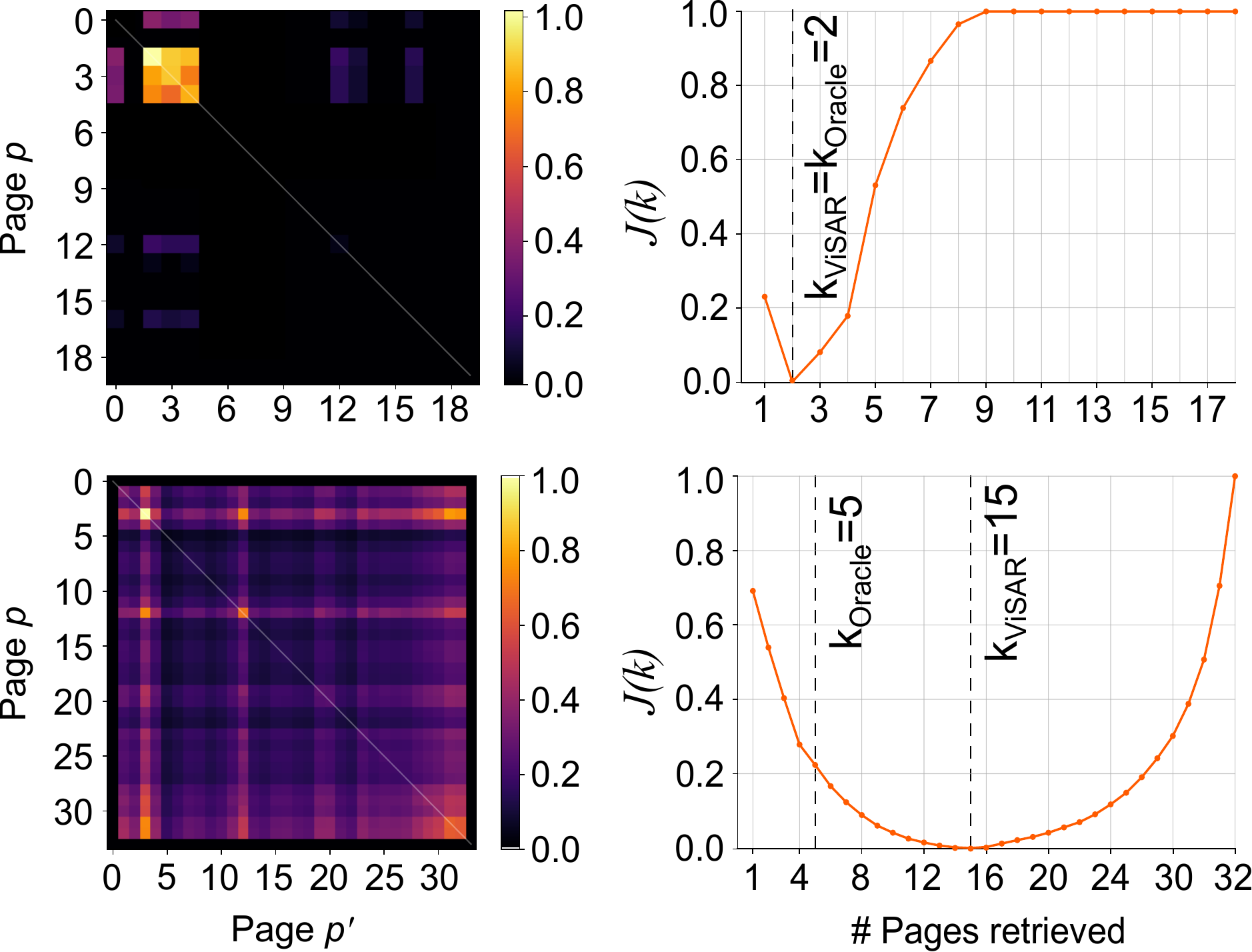}\\
		\caption{Query-conditioned similarity matrices (zero values in black) and the corresponding retrieval cost function $\mathcal{J}(k)$, with $k^\star_{\text{ViSAR}}$ and $k^\star_{\text{Oracle}}$ marked. \textbf{Upper}: A sparse matrix allows effective localization of query semantics, which produces a sharp minimum in $\mathcal{J}(k)$, enabling accurate adaptive retrieval. \textbf{Lower}: A dense matrix yields a shallow minimum, causing over-retrieval.\label{fig:curve_heatmap}}
		
		\vspace{1em}
		
		\includegraphics[width=0.9\linewidth]{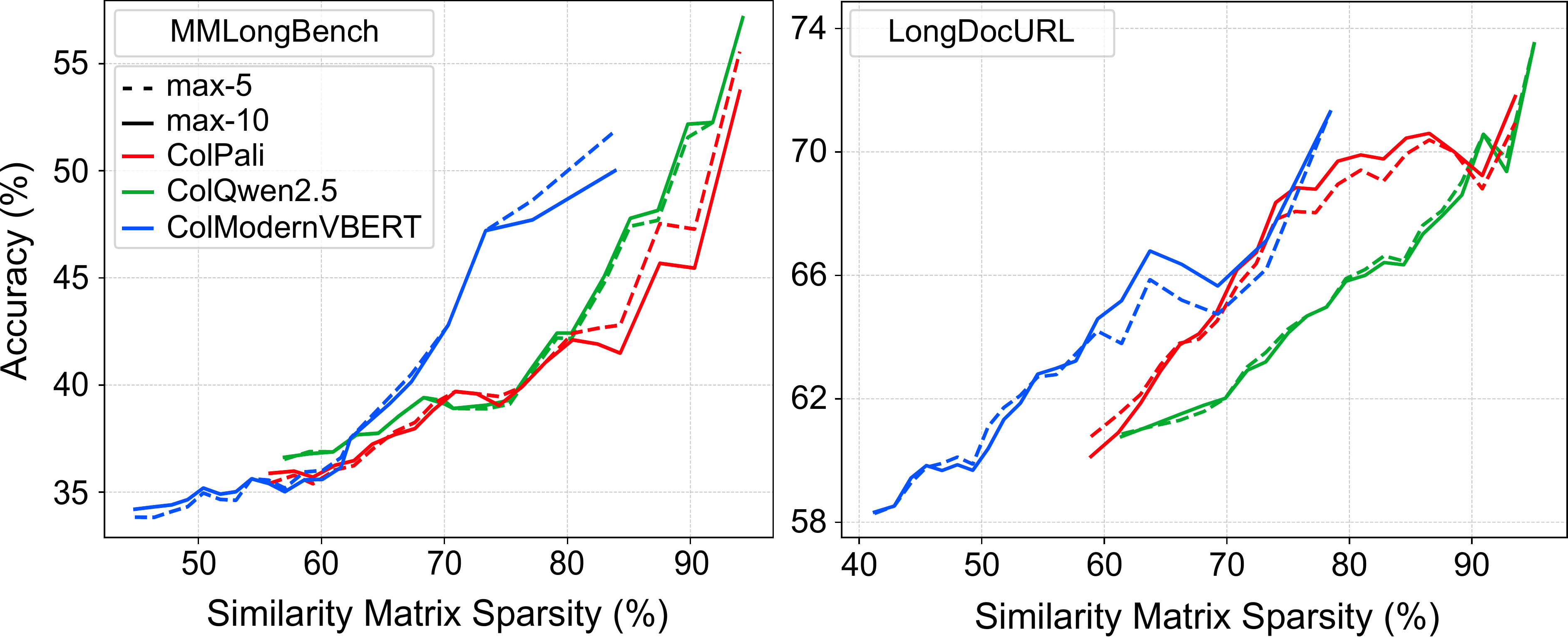}\\
		\caption{Query-conditioned similarity matrix sparsity, showing answer accuracy as a function of the percentage of inactive entries in the similarity matrix, using Qwen2.5-VL-7B-Instruct. Higher sparsity consistently correlates with higher accuracy across encoders, budgets, and datasets.\label{fig:accuracy_vs_simpp_sparsity}}
	\end{figure}
	
	ViSAR's query-conditioned similarity matrix $\mathrm{Sim}(p,p')$ provides an explicit representation of how query-relevant information is distributed across a document. \autoref{fig:curve_heatmap} illustrates representative examples and shows that, when query semantics are localized, $\mathrm{Sim}(p,p')$ is sparse and produces a sharp minimum in $\mathcal{J}(k)$, enabling a reliable retrieval boundary. Conversely, when query semantics are distributed across many pages, $\mathrm{Sim}(p,p')$ becomes denser, yielding a shallow minimum and a less reliable stopping decision.
	
	\autoref{fig:accuracy_vs_simpp_sparsity} confirms this trend, showing that higher similarity matrix sparsity consistently correlates with higher answer accuracy across encoders, budgets, and datasets. While $\mathcal{J}(k)$ is designed to maximise coherence and minimise leakage within this structure, these results provide post-hoc evidence that its components capture semantically meaningful retrieval properties.
	
	\subsubsection{Ablation Study}
	
	\begin{table}[t!]
		\begin{minipage}{\linewidth}
			\centering
			\small
			\setlength{\tabcolsep}{1mm}
			\caption{Ablation study on page ranking, reporting NDCG and recall at ranks 5 and 10. A check mark indicates that the weight is applied and a blank indicates that it is set to 1. Removing any weighting component degrades retrieval quality, showing that each component contributes to ViSAR's performance. Setting $w^{p}_{j}=1$ makes all self-similarities $s_p=\mathrm{Sim}(p,p)$ equal to 1, so pages cannot be ranked.\label{tab:ablation_weights}}	
			\begin{tabularx}{\linewidth}{
					c c c
					*{8}{>{\centering\arraybackslash}X}
				}
				\toprule
				
				\multicolumn{3}{c}{\textbf{Weights}} & \multicolumn{4}{c}{\textbf{MMLongBench}} & \multicolumn{4}{c}{\textbf{LongDocURL}} \\
				\cmidrule(lr){1-3} \cmidrule(lr){4-7} \cmidrule(lr){8-11}
				
				$w_i$ & $w_p$ & $w^{p}_{j}$ 
				& NDCG@5 & NDCG@10 & Recall@5 & Recall@10 
				& NDCG@5 & NDCG@10 & Recall@5 & Recall@10 \\
				
				\midrule
				
				\checkmark & \checkmark & \checkmark
				& 0.734          & \textbf{0.756} & \textbf{76.36} & \textbf{86.68}
				& \textbf{0.790} & \textbf{0.798} & \textbf{76.44} & \textbf{84.44} \\
				
				& \checkmark & \checkmark
				& \textbf{0.735} & 0.748          & 75.04          & 85.15          
				& 0.772          & 0.779          & 74.52          & 83.08          \\
				
				\checkmark & & \checkmark
				& 0.684          & 0.702          & 70.48          & 81.50
				& 0.716          & 0.730          & 71.07          & 80.45          \\
				
				&  & \checkmark
				& 0.615          & 0.646          & 66.09          & 80.15
				& 0.561          & 0.594          & 58.29          & 72.50          \\
				
				\checkmark & \checkmark &
				& \multicolumn{4}{c}{Degenerate}
				& \multicolumn{4}{c}{Degenerate} \\
				
				\bottomrule
			\end{tabularx}
		\end{minipage}
		
		\vspace{1em}
		
		\begin{minipage}{0.48\linewidth}
			\centering
			
			\small
			\caption{Ablation study on the retrieval objective, reporting answer accuracy under a Max-10 budget. $\mathcal{J}(k)$ is replaced with fixed top-$k$, Largest-Gap, or Score-Cluster selection. The rest of the ViSAR pipeline is unchanged. Replacing $\mathcal{J}(k)$ lowers accuracy in all settings.\label{tab:ablation_objective}}
			\setlength{\tabcolsep}{0.4mm}
			
			\begin{tabularx}{\linewidth}{
					>{\raggedright\arraybackslash}p{0.26\linewidth}
					*{4}{>{\centering\arraybackslash}X}
				}
				\toprule
				& \multicolumn{2}{c}{\textbf{MMLongBench}} & \multicolumn{2}{c}{\textbf{LongDocURL}} \\
				\cmidrule(lr){2-3} \cmidrule(lr){4-5}
				\textbf{Variant} & Mean $k^\star$ & Max-10 & Mean $k^\star$ & Max-10 \\
				\midrule
				
				ViSAR          & 5.3  & \textbf{35.88} & 8.1   & \textbf{60.13} \\
				Fixed top-$10$ & --   & 35.78          & --   & 58.84          \\			
				Largest-Gap    & 1.7  & 33.55          & 2.2  & 59.36          \\
				Score-Cluster  & 11.6 & 35.14          & 22.6 & 59.18          \\
				
				\bottomrule
			\end{tabularx}
		\end{minipage}	
		\hfil
		\begin{minipage}{0.48\linewidth}
			\centering
			\includegraphics[width=\linewidth]{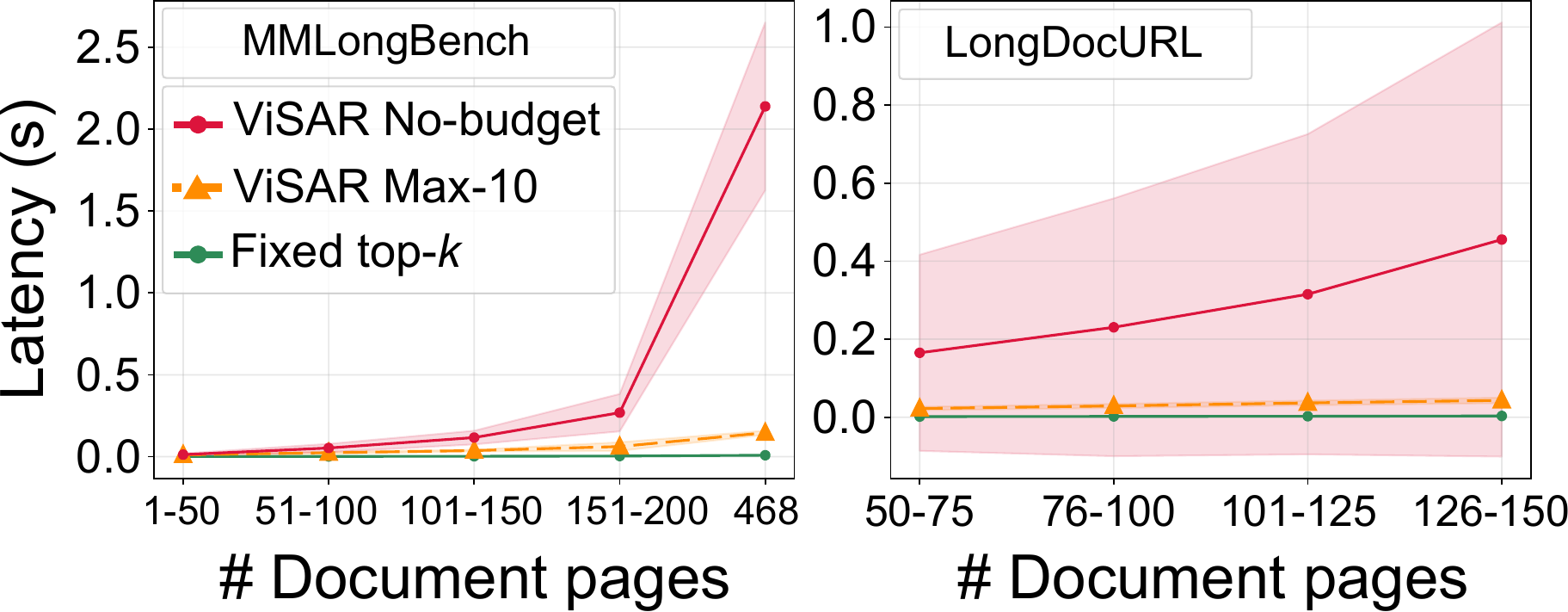} \\
			\figurecaption{Ablation study on the budget, reporting latency under no budget (i.e, $\text{Max-}k=N$), which computes the full similarity matrix and evaluates $\mathcal{J}(k)$ for all $k$. Without a budget constraint, the retrieval latency of ViSAR increases substantially.\label{fig:ablation_budget_latency}}
			\vfill
		\end{minipage}
	\end{table}
	
	\autoref{tab:ablation_weights} reports the ablation of the weighting strategy, setting the query embedding weights $w_i$, the page weights $w_p$, or both to 1. Removing either weight degrades page ranking, and removing both causes the largest drop in all metrics, showing that the two weights are complementary. In addition, setting the patch weights $w^{p}_{j}$ to 1 makes all self-similarities $s_p=\mathrm{Sim}(p,p)$ equal to 1, which prevents page ranking. 
	
	\autoref{tab:ablation_objective} reports the ablation of the retrieval objective, replacing $\mathcal{J}(k)$ with fixed top-$k$, Largest-Gap, or Score-Cluster selection on top of the ViSAR pipeline. All replacements reduce accuracy, with Largest-Gap under-retrieving (mean $k^\star$ of 1.7 and 2.2) and Score-Cluster over-retrieving (mean $k^\star$ of 11.6 and 22.6). Overall, both the weighting strategy and the retrieval objective contribute to ViSAR's performance.
	
	\autoref{fig:ablation_budget_latency} reports the ablation of the budgeted constraint, obtained by setting $\text{Max-}k=N$, which computes the full similarity matrix and evaluates $\mathcal{J}(k)$ for all $k$. Without the budget, the retrieval latency of ViSAR increases substantially, with the largest increase observed for the longest document (468 pages, MMLongBench), from 0.15 s to 2.14 s ($\approx$14$\times$). The budget therefore contributes to the low retrieval overhead of ViSAR, without changing the pages sent to the LVLM.
	
	\section{Discussion}
	
	ViSAR reduces end-to-end RAG latency by up to 58.9\%, adapting the number of retrieved pages to each query while preserving or improving accuracy: across 60 configurations (5 LVLMs $\times$ 3 encoders $\times$ 2 budgets $\times$ 2 datasets), ViSAR yields a statistically significant accuracy improvement in 27 and shows no significant loss in the remaining 33. Its budget-aware computation effectively restricts the page-level similarity matrix to the pages that can be passed to the LVLM, and selects an adaptive-$k$ set of pages, enabling a substantial reduction in generation latency with a marginal retrieval overhead (\autoref{fig:latency_RAG_VISAR}). In addition, ViSAR hyperparameters $T$ and $\gamma$ require no per-dataset tuning. The ranking gains over late-interaction are positive over a wide range of $T$, and a single default costs little compared with the best value for each encoder, while $\gamma$ has a plateau spanning several orders of magnitude.
	
	In comparison, while Largest-Gap provides higher latency gains in LongDocURL, it reaches a higher F1-score only on single-pages queries, which dominate the dataset and therefore benefit latency. Largest-Gap also exhibits a heavy-tailed distribution, which may saturate the budget more often. Therefore, its latency advantage is expected to shrink as the budget increases. Largest-Gap also yields occasional statistically significant accuracy losses, with an accuracy at best comparable to that of ViSAR. Similar observations also apply to Score-Cluster although it shows no latency advantages. Conversely, ViSAR shows no significant accuracy loss, with significant gain in some settings, and retrieves more stable page sets across query complexities. This contrast likely stems from the simple adaptive mechanism of Largest-Gap and Score-Cluster, which only observe the one-dimensional ranked list of independent page scores from late-interaction. They are therefore expected to work best when the encoder produces a clear cut-off in the scores, whereas noisier scores from complex queries may explain their tendency to retrieve substantially more pages than required. ViSAR instead builds a query-conditioned similarity matrix that captures how query semantics are distributed across pages, and the cost function $\mathcal{J}(k)$ exploits this structure to select a coherent page set while minimizing the leakage from the remaining pages.
	
	However, ViSAR also depends on the ability of the encoders to represent documents and queries: underspecified queries or homogeneous documents may increase the difficulties to obtain discriminative semantics, resulting in more leakage and limiting the possibility to retrieve a compact page set. This is reflected in the structure of the similarity matrix, where more leakage is expected to decrease its sparsity, which is associated with decreasing accuracy. The encoders capability may also partly explain the tendency of ViSAR to inflate the average number of retrieved pages $k^\star$ as document length increases: additional pages may introduce more leakage, requiring more pages to be selected to minimize the leakage. Finally, it also implies that part of the difficulty lies in the ranking and not only in the retrieval mechanism, as shown by the behaviour of the Oracle that often needs to retrieve significantly more pages than the actual number of evidence pages.
	
	Yet, ViSAR still achieves an effective adaptive-$k$ retrieval in most cases, while allowing the localization of query semantics across the documents. Therefore, future investigation should further explore the structure of the similarity matrix to explain why some queries fail to yield compact page sets, which Largest-Gap and Score-Cluster cannot allow by only using the one-dimensional list of page scores. Future investigation could also further reduce the average number of retrieved pages, possibly reducing the latency beyond what Largest-Gap achieves on LongDocURL. Furthermore, a promising direction could be to understand the conditions of success and failure, potentially enabling the design of a label-free feedback signal, for iterative strategies when single-pass retrieval is insufficient: methods such as multi-round query refinement or evidence selection have the potential to overcome RAG limitations, such as poor ranking quality or the need to reason over a larger number of evidence pages, which could be guided by the structure of ViSAR's similarity matrix, without additional models or training.
	
	\section{Conclusion}
	
	We introduced ViSAR, a training-free, budget-aware visual embedding weighting mechanism enabling adaptive-$k$ retrieval for document visual question answering. It operates in the embedding space of visual late-interaction encoders to construct a query-conditioned page-level similarity matrix that drives compact and adaptive page selection, reducing RAG latency by up to 58.9\% while preserving or improving answer accuracy. We also showed that the structure of the similarity matrix reflects the localization of query-relevant information and correlates with answer accuracy, suggesting that its structure may provide a useful feedback signal for future retrieval strategies.
	
	\section{Supporting Information}
	
	Algorithm S1, Figure S1--S9, and Table S1--S2 are available in the supplementary material pdf.

	\bibliographystyle{unsrtnat}
	\bibliography{references}
	
	\appendix

\renewcommand{\thefigure}{S\arabic{figure}}
\renewcommand{\thetable}{S\arabic{table}}
\renewcommand{\thealgorithm}{S\arabic{algorithm}}

\setcounter{figure}{0}
\setcounter{table}{0}
\setcounter{algorithm}{0}

\renewcommand{\theHfigure}{S\arabic{figure}}
\renewcommand{\theHtable}{S\arabic{table}}
\renewcommand{\theHalgorithm}{S\arabic{algorithm}}

\clearpage
\newpage
\section{ViSAR Implementation Details} \label{supplementals}

\subsection{ViSAR Algorithm}

\begin{algorithm}[h!]
	\caption{ViSAR: Adaptive-$k$ Retrieval via Multi-Level Interaction Weighting}
	\label{alg:visar}
	\begin{algorithmic}[1]
		\Require $Q = \{q_i\}_{i=1}^{m}$; $D = \{P^p\}_{p=1}^{N}$; $P^p = \{v^p_j\}_{j=1}^{n_p}$; $T$; $\gamma$; Max-$k$
		\Statex
		\State \textcolor{stagecolor}{\textbf{Stage 1 -- Query-to-Page Weighting}}
		\For{each page $p$, query embedding $q_i$}
		\State $A_{p,i} \gets \max_{j} \langle q_i, v^p_j \rangle$ \Comment{Eq.~2}
		\EndFor
		\State Compute $\hat{A}_{p,i}$, $\hat{\sigma}_i$, $\tilde{A}_{p,i}$, $w_i$, $C^p_{i,i'}$, and $w_p$ \Comment{Eq.~3--8}
		\State \textcolor{stagecolor}{\textbf{Stage 2 -- Page-to-Query Weighting}}
		\State Min-Max normalize $w_i \to \hat{w}_i$, $w_p \to \hat{w}_p$
		\For{each page $p$}
		\For{each patch embedding $v^p_j$}
		\State $r^{p}_{j} \gets \max_{i} \Big[ \langle v^{p}_j,\; q_i \rangle \cdot \big( \tilde{A}_{p,i} \cdot \hat{w}_{i} \cdot \hat{w}_{p} \big)^2 \Big]$ \Comment{Eq.~9}
		\EndFor
		\EndFor
		\State $w^p_j \gets \max\big(0,\, r^p_j - \mathrm{mean}_{p,j}(r^p_j)\big)$ \Comment{Eq.~10}
		\State \textcolor{stagecolor}{\textbf{Stage 3 -- Page-to-Page Similarity}}
		\State Min-Max normalize $w^p_j \to \hat{w}^p_j$
		\For{each page $p$}
		\State $s_p \gets \mathrm{Sim}(p,p)$ \Comment{Eq.~11--12}
		\EndFor
		\State $K \gets \min(\text{Max-}k,\, N-1)$
		\State $\mathcal{P}_{\mathrm{cand}} \gets$ top-$K$ pages by $s_p$
		\For{each source page $p \in \mathcal{P}_{\mathrm{cand}}$}
		\For{each target page $p'$}
		\If{$p' \neq p$} \Comment{Diagonal similarity already computed}
		\For{each source patch $v^p_j$}
		\State $S^{p \to p'}_j \gets \hat{w}^p_j \cdot \max_{j'} \left[ \langle v^p_j, v^{p'}_{j'} \rangle \cdot \hat{w}^{p'}_{j'} \right]$ \Comment{Eq.~11}
		\EndFor
		\State $\mathrm{Sim}(p, p') \gets \sqrt{\frac{1}{T} \sum_{j \in \mathcal{T}} S^{p \to p'}_j}$ \Comment{Eq.~12, top-$T$ patches}
		\EndIf
		\EndFor
		\EndFor
		\State \textcolor{stagecolor}{\textbf{Stage 4 -- Adaptive-$k$ Retrieval}}
		\For{$k = 1$ to $K$}
		\State $\mathcal{R}_k \gets$ top-$k$ pages by $s_p$
		\State $\mathcal{I}_k \gets N-k$ remaining pages
		\State Compute $c_k^p$ and $l_k^p$ \Comment{Eq.~13--14}
		\State $\mathcal{J}(k) \gets \sum_{p \in \mathcal{R}_k} w_{s_p} \left( c^p_k - \gamma\, l^p_k \right)$ \Comment{Eq.~15}
		\EndFor
		\State $k^\star \gets \arg\min_{1 \leq k \leq K} \mathcal{J}(k)$ \Comment{Eq.~16}
		\If{$k^\star + 1 \leq K$}
		\State $n \gets \min\big(\operatorname{round}(\ln(k^\star)),\, K-k^\star\big)$ \Comment{Limit page inclusion when $k^\star$ is big}
		\State $v_{+} \gets \sum_{i=0}^{n-1} |\mathcal{J}(k^\star+i)-\mathcal{J}(k^\star+i+1)|$ \Comment{variation beyond $k^\star$}
		\State $v_{-} \gets \sum_{i=0}^{n-1} |\mathcal{J}(k^\star-i)-\mathcal{J}(k^\star-i-1)|$ \Comment{variation before $k^\star$}
		\If{$v_{+} > v_{-}$}
		\State $k^\star \gets k^\star + 1$
		\EndIf
		\EndIf
		\State \Return $\mathcal{R}_{k^\star}$
	\end{algorithmic}
\end{algorithm}

\FloatBarrier

\clearpage
\newpage
\subsection{Computational Resources}

\begin{itemize} \label{sec:ressources}
	\item \textbf{Hardware.} Experiments were conducted on a workstation equipped with:
	\begin{itemize}
		\item NVIDIA RTX A6000 GPU (48 GB VRAM, driver 560.35.03);
		\item two Intel Xeon Silver 4210R CPUs (20 physical cores, 40 threads);
		\item 251 GB of system memory.
	\end{itemize}
	
	\item \textbf{Software.} Experiments were performed on Debian GNU/Linux 12 (Bookworm) using:
	\begin{itemize}
		\item Python 3.12.9;
		\item PyTorch 2.4.0 with CUDA 12.1;
		\item Transformers 4.57.0;
		\item FlashAttention 2.7.4.post1;
		\item ColPali Engine 0.3.13.
		\item Tesseract 5.5.2 with pytesseract 0.3.13
	\end{itemize}
	The complete list of Python dependencies and their versions is provided in the released code repository (\texttt{requirements.txt}).
	
	\item \textbf{Models.} Experiments were conducted using publicly available models, identified via Hugging Face model names or GitHub repository links as listed below:
	\begin{itemize}
		\item \textbf{Retrieval -- ViSAR's encoders:}
		\begin{itemize}
			\item \texttt{vidore/colpali-v1.2};
			\item \texttt{vidore/colqwen2.5-v0.1};
			\item \texttt{ModernVBERT/colmodernvbert}.
		\end{itemize}
		\item \textbf{Retrieval -- Reproduced baselines:}
		\begin{itemize}
			\item \texttt{openbmb/VisRAG-Ret}.
			\item \texttt{https://github.com/bloomberg/m3docrag}.
			\item \texttt{colbert-ir/colbertv2.0},  using \texttt{https://github.com/AnswerDotAI/RAGatouille}.
		\end{itemize}
		\item \textbf{Retrieval -- Adapted baselines:}
		\begin{itemize}
			\item Largest-Gap:
			\begin{itemize}
				\item Adapted from: \emph{Taguchi, C.; Maekawa, S.; and Bhutani, N. 2025. Efficient Context Selection for Long-Context QA: No Tuning, No Iteration, Just Adaptive-k. In Proceedings of the 2025 Conference on Empirical Methods in Natural Language Processing, 20116–20141.}
				\item Implementation in: \texttt{./pipeline/retrievers.py::HeuristicsSelector.LargestGap}
			\end{itemize}
			\item Score-Cluster: 
			\begin{itemize}
				\item Adapted from: \emph{Xu, Y.; Gupta, V.; Aggarwal, R.; Mahadevan, V.; and Krishnamachari, B. 2025. Cluster-based Adaptive Retrieval: Dynamic Context Selection for RAG Applications. arXiv preprint arXiv:2511.14769.}
				\item Implementation in: \texttt{./pipeline/retrievers.py::HeuristicsSelector.ScoreCluster}
			\end{itemize}
		\end{itemize}
		\item \textbf{Answer generation:}
		\begin{itemize}
			\item \texttt{Qwen/Qwen2.5-VL-7B-Instruct};
			\item \texttt{Qwen/Qwen3-VL-8B-Instruct};
			\item \texttt{OpenGVLab/InternVL3\_5-8B-Instruct};
			\item \texttt{HuggingFaceM4/Idefics3-8B-Llama3};
			\item \texttt{llava-hf/llava-onevision-qwen2-7b-ov-hf};
			\item \texttt{Qwen/Qwen2.5-7B-Instruct}.
		\end{itemize}
		\item \textbf{Evaluation:}
		\begin{itemize}
			\item \texttt{Qwen/Qwen2.5-14B-Instruct}.
		\end{itemize}			
	\end{itemize}
\end{itemize}

\FloatBarrier

\clearpage
\newpage
\section{ViSAR Adaptive-$k$ Retrieval}

\subsection{Retrieved Page Numbers Distribution.}

\begin{figure}[h!]
	\centering
	\includegraphics[width=\linewidth]{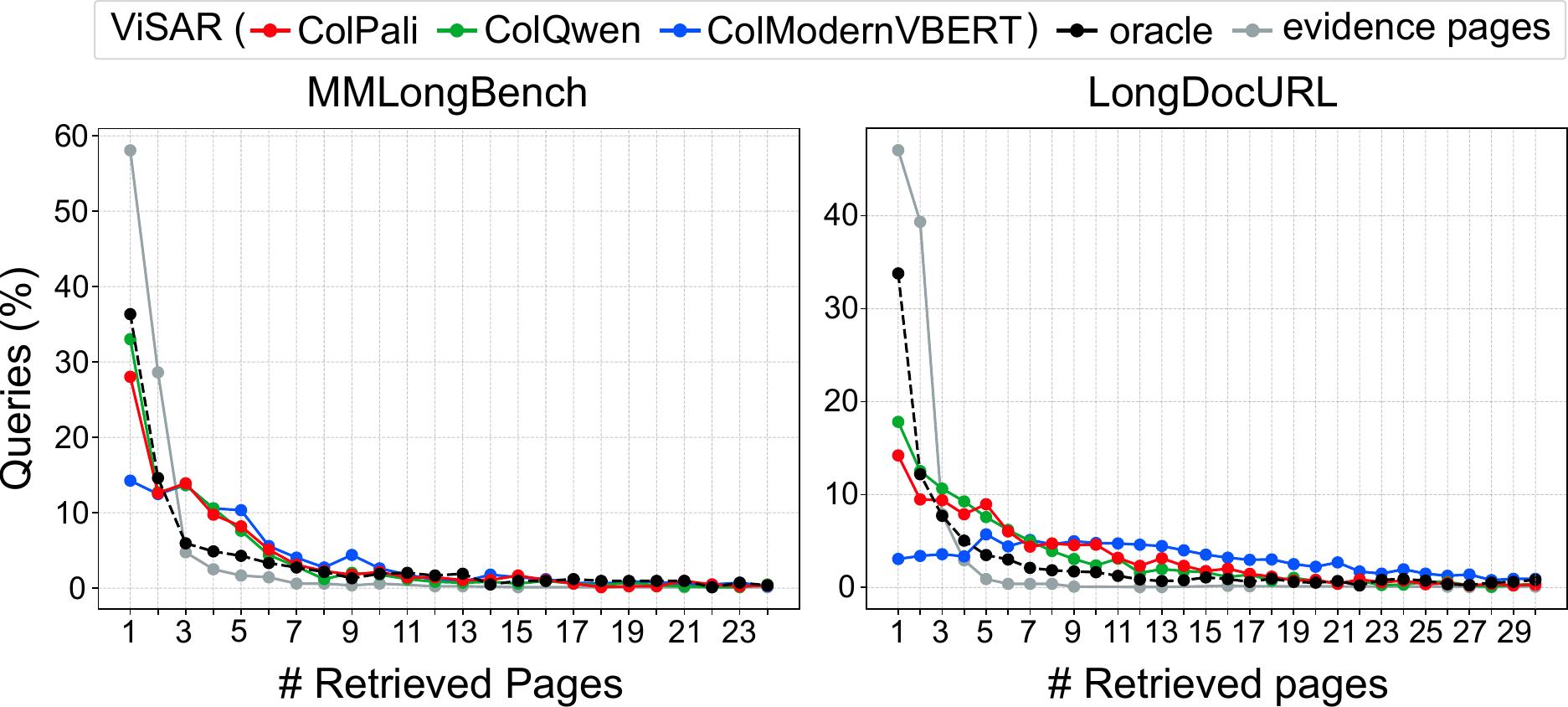}\\
	\caption{\textbf{Number of retrieved pages distribution}. ViSAR and Oracle, along with evidence pages. They show an exponentially decaying distribution (capped at the maximum evidence pages size for clarity -- statistics are reported in Table~1 of the main text). ColPali and ColQwen2.5 show similar trends to Oracle. ColModernVBERT shows a weaker trend, especially on LongDocURL.}
	\label{fig:distributions_k_star_ViSAR_Oracle}
\end{figure}

\begin{figure}[h!]
	\centering
	\includegraphics[width=\linewidth]{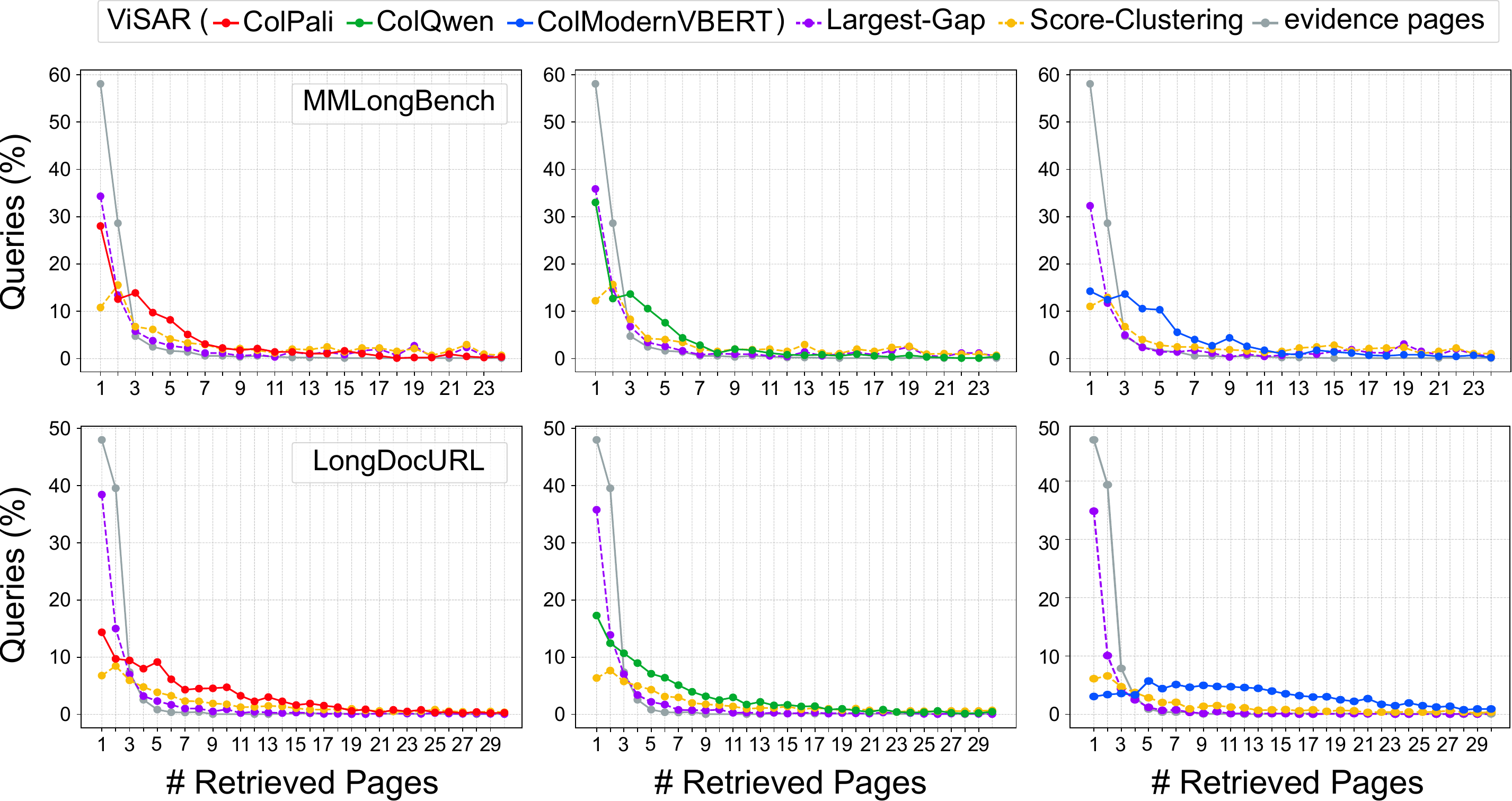}\\
	\caption{\textbf{Number of retrieved pages distribution}. ViSAR, Largest-Gap, and Score-Cluster methods, along with evidence pages. \textit{Score-Cluster}: identifies an adaptive cut-off by clustering the late-interaction scores. \textit{Largest-Gap}: uses the largest gap in the late-interaction scores as an adaptive cut-off. All methods show an exponentially decaying distribution (capped at the maximum evidence page set size for clarity; statistics are reported in Table~1 of the main text). ViSAR tends to retrieve compact sets, favouring lower and more varied page counts. In contrast, Largest-Gap tends to favour single page retrieval, while Score-Cluster shows a weaker trend overall.}
	\label{fig:distributions_k_star_heuristic_methods}
\end{figure}

\FloatBarrier

\clearpage
\newpage
\subsection{Retrieval Quality vs Oracle: Mean Difference in Retrieved size.}

\begin{figure}[th!]
	\centering
	\includegraphics[width=\linewidth]{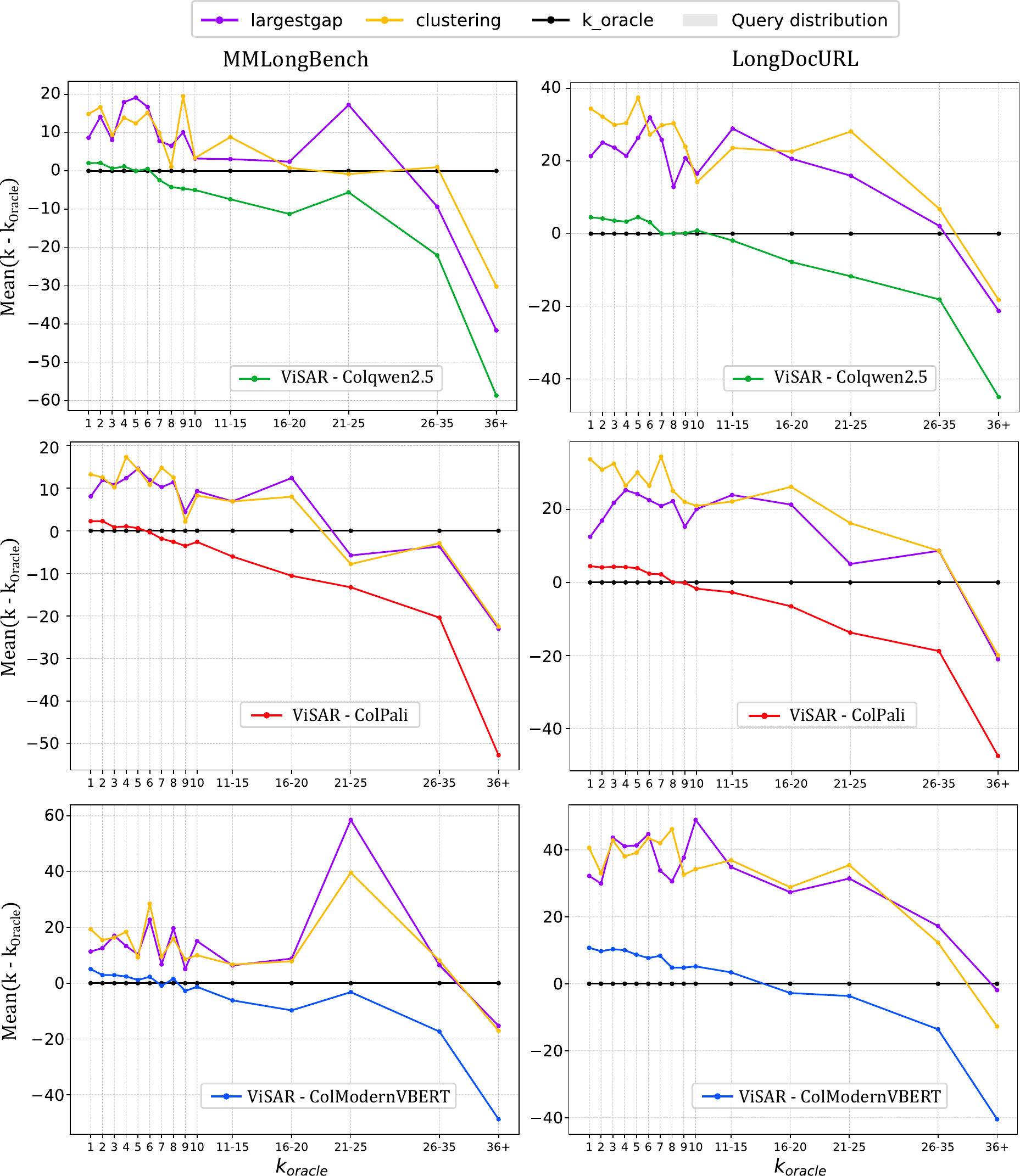}\\
	\caption{\textbf{Mean difference of $k-k_{Oracle}$ as a function of $k_{Oracle}$.} ViSAR closely follows Oracle at low $k_{\text{Oracle}}$ and retrieves fewer pages as $k_{\text{Oracle}}$ increases, limiting the retrieval of irrelevant pages. Largest-Gap and Score-Cluster over-retrieve across most $k_{\text{Oracle}}$. The trend is consistent across all three encoders and two datasets.}
	\label{fig:Metric_vs_k_oracle_k_diffs}
\end{figure}

\FloatBarrier

\clearpage
\newpage	
\subsection{Retrieval Quality vs Oracle: F1-score.}

\begin{figure}[h!]
	\centering
	\includegraphics[width=\linewidth]{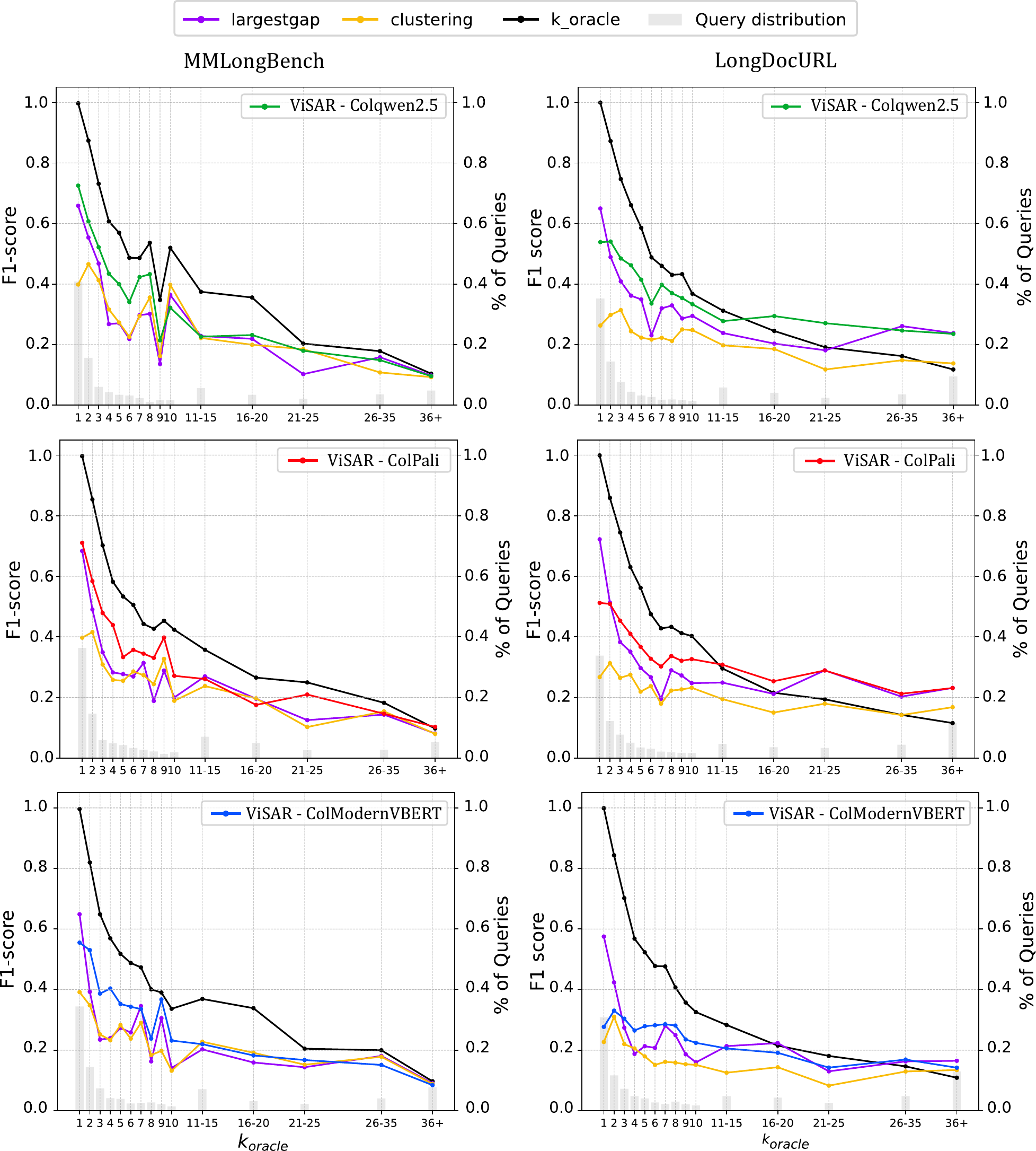}\\
	\caption{\textbf{F1-score as a function of $k_{Oracle}$.} Gray histograms show the corresponding percentage of queries. ColQwen2.5 shows the best results, followed by ColPali, while ColModernVBERT exhibits a weaker trend. ViSAR achieves a higher F1-score on MMLongBench, particularly for lower $k_{\text{Oracle}}$ where most queries are located. On LongDocURL, ViSAR outperforms Largest-Gap for most $k_{\text{Oracle}}$, except $k_{\text{Oracle}}=1$, suggesting that Largest-Gap is mostly effective for single-page queries, which constitute the majority of queries. Overall, ViSAR provides a more consistent adaptive retrieval, in accordance with its compact but more varied retrieved set size in Figure~S1 and S2}
	\label{fig:Metric_vs_k_oracle_F1-score}
\end{figure}	

\FloatBarrier

\clearpage
\newpage
\subsection{Retrieval Quality vs Oracle: Precision.}

\begin{figure}[h!]
	\centering
	\includegraphics[width=\linewidth]{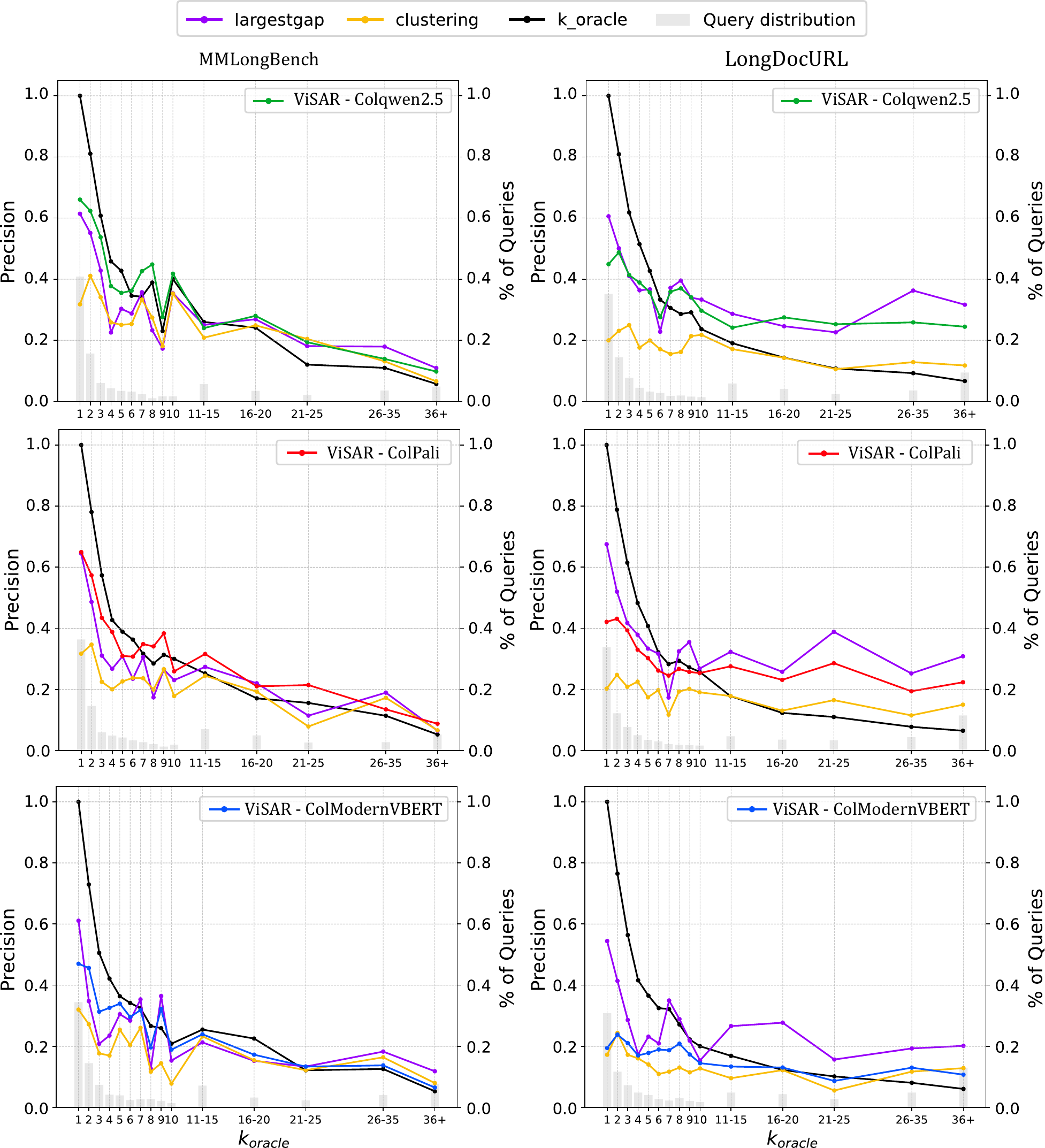}\\
	\caption{\textbf{Precision as a function of $k_{Oracle}$.} Gray histograms show the corresponding percentage of queries. ColQwen2.5 shows the best results, followed by ColPali, while ColModernVBERT exhibits a weaker trend. ViSAR achieves a higher precision on MMLongBench, particularly for lower $k_{\text{Oracle}}$ where most queries are located. On LongDocURL, when using ColQwen2.5, ViSAR and Largest-Gap achieve competitive performance on most $k_{\text{Oracle}}$, except $k_{\text{Oracle}}=1$, while ColPali and ColModernVBERT favour Largest-Gap.}
	\label{fig:Metric_vs_k_oracle_Precision}
\end{figure}

\FloatBarrier

\clearpage
\newpage
\subsection{Retrieval Quality vs Oracle: Recall.}	

\begin{figure}[h!]
	\centering
	\includegraphics[width=\linewidth]{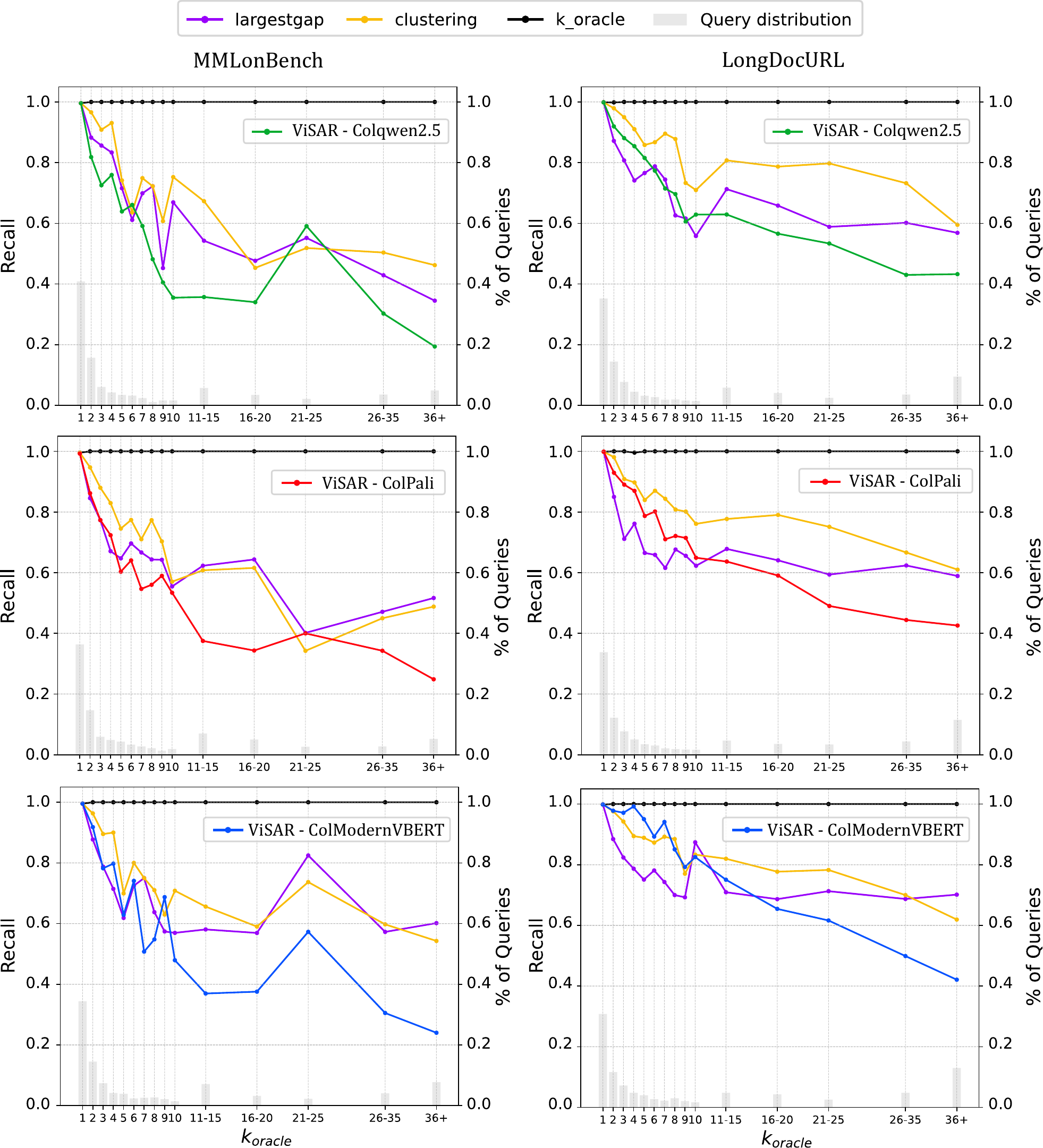}\\
	\caption{\textbf{Precision as a function of $k_{Oracle}$.} Gray histograms show the corresponding percentage of queries. ColQwen2.5 shows the best results, followed by ColPali, while ColModernVBERT exhibits a weaker trend. While Oracle always maximises Recall, by definition, adaptive methods favour lower numbers of retrieved pages on average. They reduce the number of irrelevant pages as $k_{Oracle}$ increases, with Largest-Gap and Score-Cluster shows higher precision overall than ViSAR, consistent with their larger numbers of retrieved pages at each $k_{Oracle}$ in Figure~S3.}
	\label{fig:Metric_vs_k_oracle_Recall}
\end{figure}	

\FloatBarrier

\clearpage
\newpage	
\subsection{Ranking Quality Metrics}

\begin{table}[h!]
	\centering
	\setlength{\tabcolsep}{1mm}
	\caption{Page ranking evaluation on MMLongBench and LongDocURL, comparing ViSAR and late-interaction at ranks 5 and 10, reporting Recall, normalized discounted cumulative gain (NDCG), hit rate (Hits), and mean reciprocal rank (MRR). ViSAR consistently improves page ranking, contributing to the effectiveness of its adaptive retrieval behaviour. ColQwen2.5 shows the best results, followed by ColPali, while ColModernVBERT exhibits more modest improvements. \label{tab:ranking_results_both_datasets}}
	
	\textbf{(a) MMLongBench}
	
	\vspace{2mm}
	
	\begin{tabularx}{\textwidth}{
			>{\raggedright\arraybackslash}p{0.22\textwidth}
			*{2}{>{\centering\arraybackslash}X}
			*{2}{>{\centering\arraybackslash}X}
			*{2}{>{\centering\arraybackslash}X}
			*{2}{>{\centering\arraybackslash}X}
		}
		\toprule
		
		& \multicolumn{2}{c}{Recall (\%)} & \multicolumn{2}{c}{NDCG (\%)} & \multicolumn{2}{c}{Hits} & \multicolumn{2}{c}{MRR} \\
		
		\cmidrule(lr){2-3}
		\cmidrule(lr){4-5}
		\cmidrule(lr){6-7}
		\cmidrule(lr){8-9}
		
		\textbf{Ranking Method} & @5 & @10 & @5 & @10 & @5 & @10 & @5 & @10 \\
		
		\midrule
		\multicolumn{8}{l}{\textit{Using ColPali}} \\
		\midrule
		
		\hspace{5pt} Late-Interaction      & 75.00          & 84.49          & 0.730          & 0.746          & 84.92          & 90.97          & 0.696          & 0.704          \\
		\hspace{5pt} ViSAR (\textbf{ours}) & \textbf{76.36} & \textbf{86.68} & \textbf{0.734} & \textbf{0.756} & \textbf{85.39} & \textbf{92.76} & \textbf{0.702} & \textbf{0.712} \\
		
		\midrule
		\multicolumn{8}{l}{\textit{Using ColQwen2.5}} \\
		\midrule
		
		\hspace{5pt} Late-Interaction      & 78.60          & 86.82          & 0.780          & 0.793          & 87.89          & 92.87          & 0.754          & 0.760          \\
		\hspace{5pt} ViSAR (\textbf{ours}) & \textbf{79.73} & \textbf{87.73} & \textbf{0.790} & \textbf{0.799} & \textbf{88.72} & \textbf{93.23} & \textbf{0.762} & \textbf{0.768} \\
		
		\midrule
		\multicolumn{8}{l}{\textit{Using ColModernVBERT}} \\
		\midrule
		
		\hspace{5pt} Late-Interaction      & 72.79          & 83.14          & 0.704          & 0.723          & 83.14          & 90.26          & 0.668          & 0.678          \\
		\hspace{5pt} ViSAR (\textbf{ours}) & \textbf{73.66} & \textbf{83.56} & \textbf{0.709} & \textbf{0.725} & \textbf{83.85} & \textbf{90.26} & \textbf{0.671} & \textbf{0.679} \\
		
		\bottomrule
	\end{tabularx}
	
	\vspace{4mm}
	
	\textbf{(b) LongDocURL}
	
	\vspace{2mm}
	
	\begin{tabularx}{\textwidth}{
			>{\raggedright\arraybackslash}p{0.22\textwidth}
			*{2}{>{\centering\arraybackslash}X}
			*{2}{>{\centering\arraybackslash}X}
			*{2}{>{\centering\arraybackslash}X}
			*{2}{>{\centering\arraybackslash}X}
		}
		\toprule
		
		& \multicolumn{2}{c}{Recall (\%)} & \multicolumn{2}{c}{NDCG (\%)} & \multicolumn{2}{c}{Hits} & \multicolumn{2}{c}{MRR} \\
		
		\cmidrule(lr){2-3}
		\cmidrule(lr){4-5}
		\cmidrule(lr){6-7}
		\cmidrule(lr){8-9}
		
		\textbf{Ranking Method} & @5 & @10 & @5 & @10 & @5 & @10 & @5 & @10 \\
		
		\midrule
		\multicolumn{8}{l}{\textit{Using ColPali}} \\
		\midrule
		
		\hspace{5pt} Late-Interaction      & 75.07          & 83.15          & 0.777          & 0.788          & 88.72          & 93.97          & 0.750          & 0.758          \\
		\hspace{5pt} ViSAR (\textbf{ours}) & \textbf{76.44} & \textbf{84.44} & \textbf{0.790} & \textbf{0.798} & \textbf{90.14} & \textbf{95.01} & \textbf{0.762} & \textbf{0.769} \\
		
		\midrule
		\multicolumn{8}{l}{\textit{Using ColQwen2.5}} \\
		\midrule
		
		\hspace{5pt} Late-Interaction      & 77.54          & 84.98          & 0.805          & 0.810          & 90.70          & 94.88          & 0.779          & 0.785          \\
		\hspace{5pt} ViSAR (\textbf{ours}) & \textbf{78.69} & \textbf{85.69} & \textbf{0.817} & \textbf{0.820} & \textbf{91.56} & \textbf{95.09} & \textbf{0.793} & \textbf{0.798} \\
		
		\midrule
		\multicolumn{8}{l}{\textit{Using ColModernVBERT}} \\
		\midrule
		
		\hspace{5pt} Late-Interaction      & 71.55          & 81.10          & 0.724          & 0.739          & 85.18          & 92.34          & 0.690          & 0.700          \\
		\hspace{5pt} ViSAR (\textbf{ours}) & \textbf{72.30} & \textbf{81.57} & \textbf{0.736} & \textbf{0.748} & \textbf{85.88} & \textbf{92.38} & \textbf{0.703} & \textbf{0.712} \\
		
		\bottomrule
	\end{tabularx}
\end{table}	

\FloatBarrier		

\clearpage
\newpage
\section{ViSAR Sensitivity Analyses}

We analyse the sensitivity of ViSAR to $T$ (Equation~11 in main text ) and  $\gamma$ (Equation~14 in main text). Figures S7 and S8 show NDCG and Recall at ranks 5 and 10 as functions of $T \in [1, 10^{3}]$. Figure S9 shows $k^\star$ as a function of $\gamma \in [1, 10^8]$.

\subsection{Ranking Sensitivity to $T$ in Equation 11}

\begin{figure}[h]
	\centering
	\includegraphics[width=\linewidth]{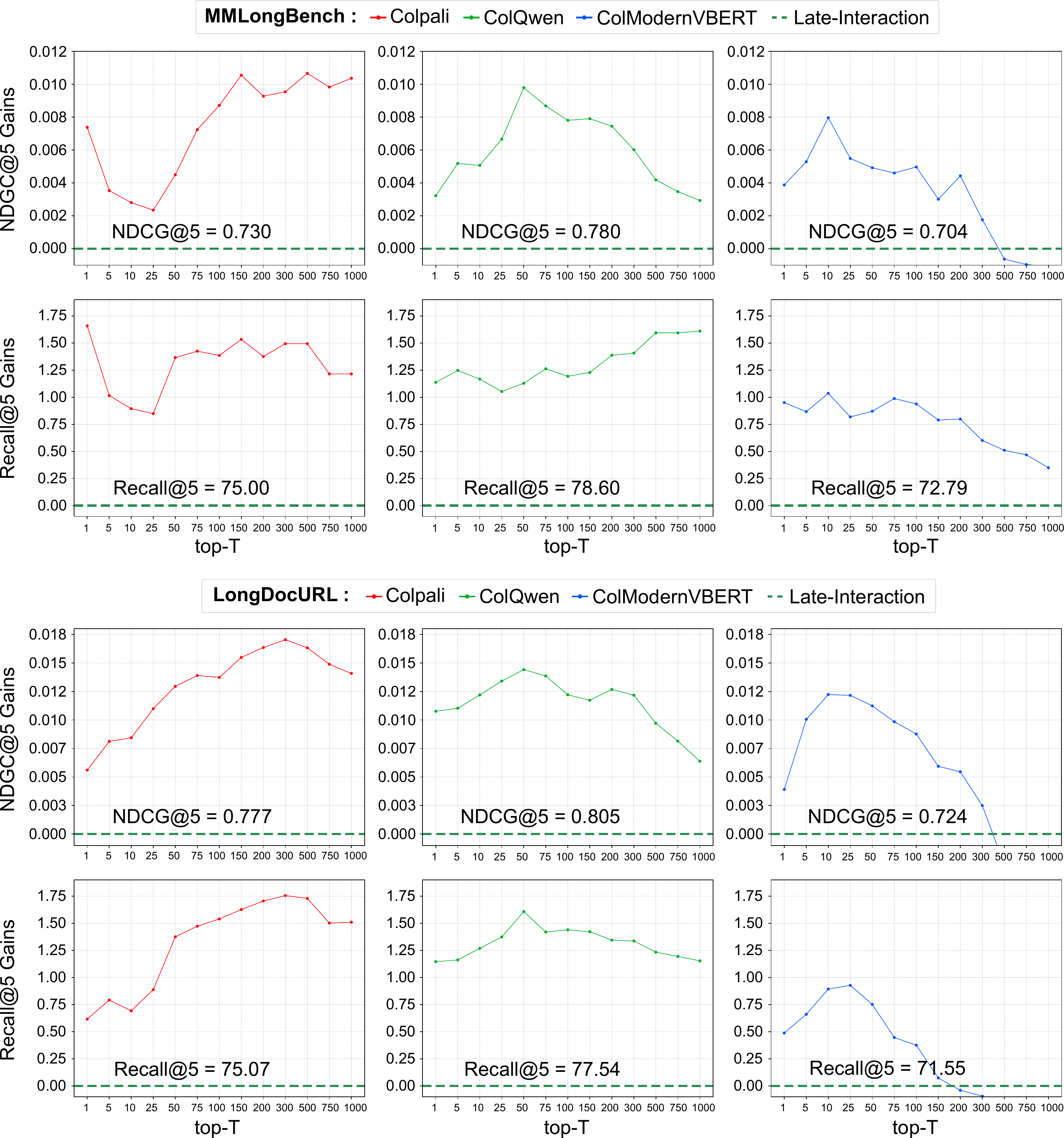}
	\caption{\textbf{Gains in NDCG@5 and Recall@5 as a function of $T \in [1,10^3]$} -- Equation 11 in the main text. Gains are relative to late-interaction (zero dashed line) and are shown for both datasets. ViSAR preserves or improves recall and NDCG for $T\leq150$ in every configuration, and for all $T$ using ColQwen2.5 and ColPali. Page ranking quality remains largely stable over a large range of $T$, with negligible variations across encoders and datasets, while enabling slight improvements.}
	\label{fig:sensitivity_at_5_top_T_k_star}
\end{figure}

\begin{figure}[h]
	\centering
	\includegraphics[width=\linewidth]{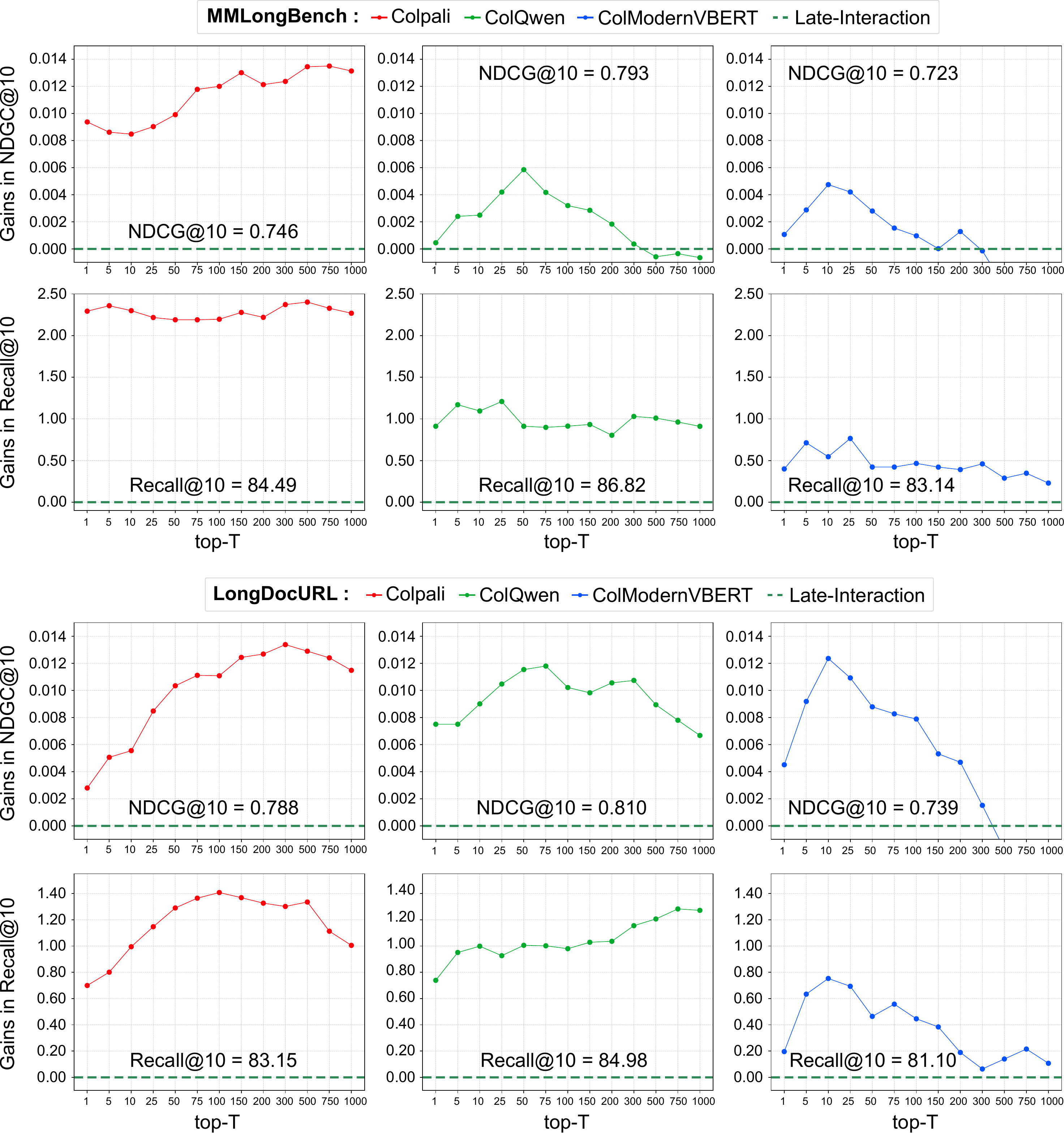}
	\caption{\textbf{Gains in NDCG@10 and Recall@10 as a function of $T \in [1,10^3]$} --Equation 11 in main text. The gains are relative to late-interaction (zero dashed line) and are shown for both datasets. ViSAR preserves or improves recall and NDCG for $T\leq150$ in every configuration. Page ranking quality remains largely stable over a large range of $T$, with negligible variations across encoders and datasets, while enabling slight improvements.}
	\label{fig:sensitivity_at_10_top_T_k_star}
\end{figure}

\FloatBarrier

\clearpage
\newpage
\subsection{Adaptive-$k$ Sensitivity to $\lambda$ in Equation 14}

\begin{figure}[h]
	\centering
	\includegraphics[width=\linewidth]{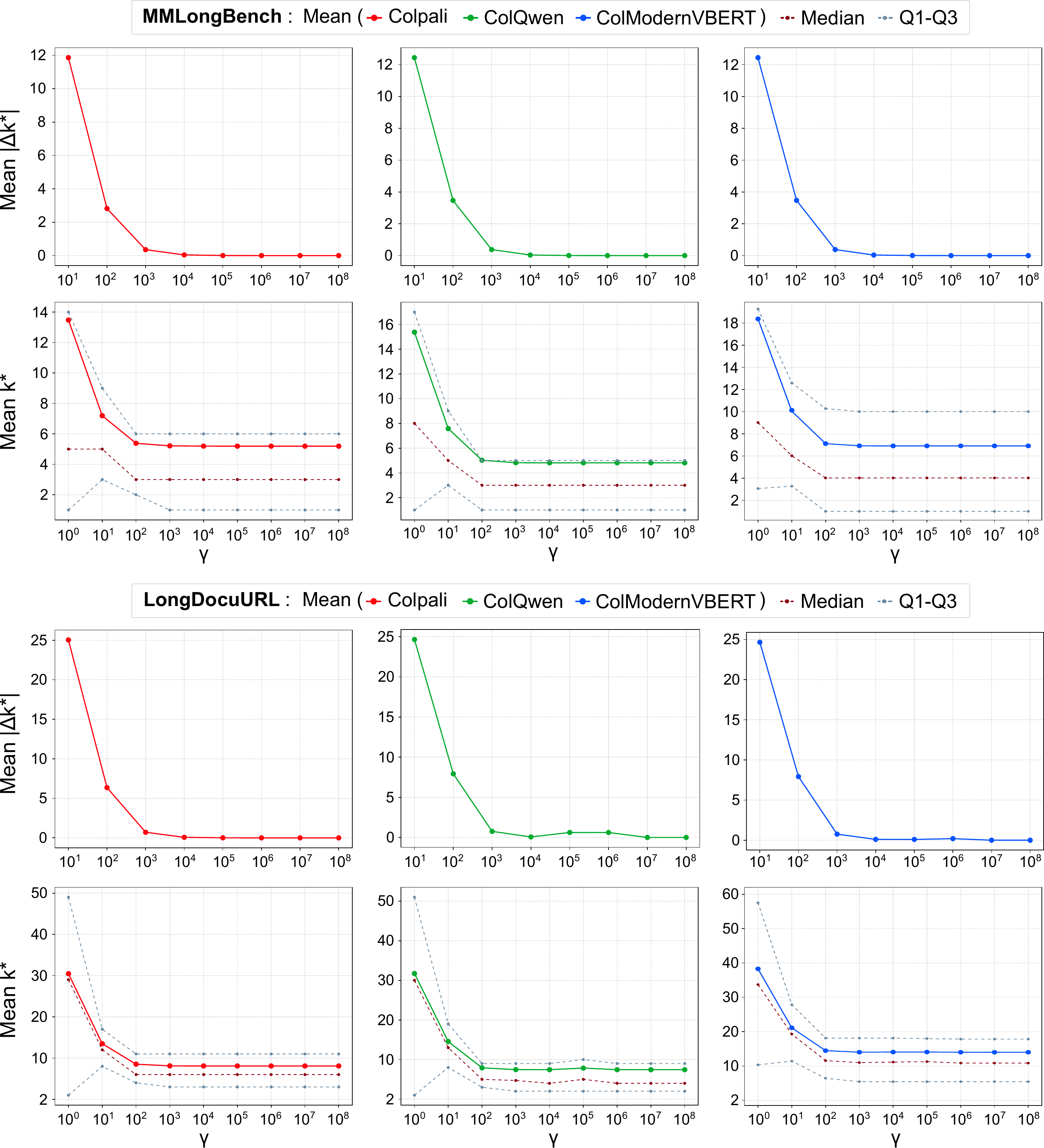}
	\caption{\textbf{$k^\star$ as a function of $\gamma \in [1,10^8]$ (Equation~14 in the main text).} Results are shown for each dataset. \textit{Top:} Mean($|\Delta k^\star|$) across queries, computed as the average absolute change in $k^\star$ between consecutive values $\gamma_i$ and $\gamma_{i-1}$, with the abscissa reporting $\gamma_i$. \textit{Bottom:} Mean($k^\star$) across queries as a function of $\gamma$. Small values of $\gamma$ tend to produce unstable and overly large values of $k^\star$. Beyond approximately $\gamma=10^4$, both Mean($|\Delta k^\star|$) and Mean($k^\star$) stabilize, \textbf{indicating that adaptive-$k$ retrieval is largely insensitive to the precise choice of $\gamma$ in this regime}.}
	\label{fig:sensitivity_analysis_k_star_function_of_gamma}
\end{figure}

\FloatBarrier

\clearpage
\newpage
\section{ViSAR RAG Latency}

\subsection{Comparison with Largest-Gap and Score-Cluster}

\begin{table}[h!]
	\centering
	\caption{\textbf{Latencies comparing ViSAR, Largest-Gap, and Score-Cluster.} Max-$k$ denotes an input budget of at most $k$ pages, where late-interaction fixed top-$k$ retrieval always maximises the LVLM input budget and adaptive methods only do so when required. Values are average latencies for retrieval, generation, and end-to-end (retrieval + generation), with gain percentages over late-interaction in parentheses. All adaptive methods significantly improved generation latency. Although ViSAR introduces a retrieval overhead, it contributes only marginally as it achieves a substantial reduction in generation latency that dominates the total cost. This results in end-to-end latency reductions of up to 58.9\% on MMLongBench and 39.2\% on LongDocURL at a Max-10 budget. In contrast, Score-Cluster end-to-end latency is consistently the slowest method, while Largest-Gap end-to-end latency provides a noticeable advantage only on LongDocURL, with gains of up to 47.2\% over fixed top-$k$ at Max-10 budget.}
	\label{tab:latency_breakdown_visar_vs_heuristics}
	
	\textbf{(a) Retrieval latency}
	
	\vspace{2mm}
	
	\begin{tabularx}{0.7\textwidth}{
			>{\raggedright\arraybackslash}p{0.20\textwidth}
			*{4}{>{\centering\arraybackslash}X}
		}
		\toprule
		
		& \multicolumn{2}{c}{\textbf{MMLongBench}} & \multicolumn{2}{c}{\textbf{LongDocURL}} \\
		\cmidrule(lr){2-3} \cmidrule(lr){4-5}

		\textbf{Method} & Max-5 & Max-10 & Max5 & Max-10 \\				
		
		\midrule
		\multicolumn{5}{l}{\textit{Fixed top-$k$}} \\
		\midrule		
		
		Late-Interaction      & 0.001 & 0.001 & 0.002 & 0.002 \\
		
		\midrule
		\multicolumn{5}{l}{\textit{Adaptive-$k$}} \\
		\midrule
		
		Score-Cluster         & 0.118 & 0.118 & 0.192 & 0.192 \\
		Largest-Gap           & 0.001 & 0.001 & 0.002 & 0.002 \\
		ViSAR (\textbf{ours}) & 0.015 & 0.018 & 0.025 & 0.030\\
		
		\bottomrule
	\end{tabularx}
	
	\vspace{4mm}
	
	\textbf{(a) Generation latency}
	
	\vspace{2mm}
	
	\begin{tabularx}{\textwidth}{
			>{\raggedright\arraybackslash}p{0.20\textwidth}
			*{4}{>{\centering\arraybackslash}X}
		}
		\toprule
		
		& \multicolumn{2}{c}{\textbf{MMLongBench}} & \multicolumn{2}{c}{\textbf{LongDocURL}} \\		
		\cmidrule(lr){2-3} \cmidrule(lr){4-5}
		
		\textbf{Method} & Max-5 & Max-10 & Max-5 & Max-10 \\
		
		\midrule
		\multicolumn{5}{l}{\textit{Fixed top-$k$}} \\
		\midrule		
		
		Late-Interaction      & 4.88 & 10.05 & 5.20 & 10.65 \\
		
		\midrule
		\multicolumn{5}{l}{\textit{Adaptive-$k$}} \\
		\midrule
		
		Score-Cluster            & 4.48 (-8.2\%)           & 6.61 (-34.2\%)          & 4.67 (-10.2\%)           & 8.17 (-23.3\%) \\						
		Largest-Gap           & 3.35 (-32.05\%)          & 5.45 (-46.04\%)          & \textbf{3.76} (-27.83\%)  & \textbf{5.63} (-47.19\%) \\
		ViSAR (\textbf{ours}) & \textbf{3.23} (-33.81\%) & \textbf{4.14} (-58.81\%) & 4.25 (-18.27\%)           & 6.47 (-39.25\%) \\
		
		\bottomrule
	\end{tabularx}
	
	\vspace{4mm}
	
	\textbf{(a) End-to-end latency}
	
	\vspace{2mm}
	
	\begin{tabularx}{\textwidth}{
			>{\raggedright\arraybackslash}p{0.20\textwidth}
			*{4}{>{\centering\arraybackslash}X}
		}
		\toprule
		
		& \multicolumn{2}{c}{\textbf{MMLongBench}} & \multicolumn{2}{c}{\textbf{LongDocURL}} \\
		\cmidrule(lr){2-3} \cmidrule(lr){4-5}
		
		\textbf{Method} & Max-5 & Max-10 & Max5 & Max-10 \\
		
		\midrule
		\multicolumn{5}{l}{\textit{Fixed top-$k$}} \\
		\midrule		
		
		Late-Interaction      & 4.93 & 10.10 & 5.21 & 10.66 \\
		
		\midrule
		\multicolumn{5}{l}{\textit{Adaptive-$k$}} \\
		\midrule
		
		Score-Cluster         & 4.60 (-6.69\%)           & 6.73 (-33.37\%)          & 4.86 (-6.72\%)           & 8.36 (-21.58\%)          \\
		Largest-Gap           & 3.35 (-32.05\%)          & 5.45 (-46.04\%)          & \textbf{3.76} (-27.83\%) & \textbf{5.63} (-47.19\%) \\
		ViSAR (\textbf{ours}) & \textbf{3.24} (-34.28\%) & \textbf{4.15} (-58.91\%) & 4.26 (-18.23\%)          & 6.48 (-39.21\%)          \\
		
		\bottomrule
	\end{tabularx}
\end{table}

\end{document}